\documentclass[journal]{IEEEtran}
\usepackage{amsmath,amsfonts}
\usepackage{algorithmic}
\usepackage{algorithm}
\usepackage{array}
\usepackage[caption=false,font=normalsize,labelfont=sf,textfont=sf]{subfig}
\usepackage{textcomp}
\usepackage{stfloats}
\usepackage{url}
\usepackage{verbatim}
\usepackage{graphicx}
\usepackage[dvipsnames]{xcolor}
\usepackage{cite}
\usepackage{amsthm}
\theoremstyle{definition}
\newtheorem{definition}{Definition}

\usepackage{tikz}
\usetikzlibrary{positioning}
\definecolor{node_inner}{HTML}{DAE8FC}
\definecolor{node_outer}{HTML}{6C8EBF}
\definecolor{silversand}{HTML}{BDC0C1}

\usepackage{booktabs}

\usepackage{comment}

\usepackage{makecell}

\begin{document}

\title{A Survey on Spatio-Temporal Knowledge Graph Models}

\author{Philipp Plamper, Hanna Köpcke, Anika Groß

\thanks{P. Plamper is with the Anhalt University of Applied Sciences, Department Computer Science and Languages, Lohmannstraße 23, 06366 Köthen, Germany. E-mail: philipp.plamper@hs-anhalt.de}
\thanks{H. Köpcke is with the University of Applied Sciences Mittweida, Faculty Applied Computer Sciences \& Biosciences, Technikumplatz 1, 09648 Mittweida, Germany. E-mail: koepcke@hs-mittweida.de }
\thanks{A. Groß is with the Anhalt University of Applied Sciences, Department Computer Science and Languages, Lohmannstraße 23, 06366 Köthen, Germany. E-mail: anika.gross@hs-anhalt.de}
}

\markboth{This work has been submitted to the IEEE for possible publication.}%
{Shell \MakeLowercase{\textit{et al.}}: A Sample Article Using IEEEtran.cls for IEEE Journals}


\maketitle

\begin{abstract}
Many complex real-world systems exhibit inherently intertwined temporal and spatial characteristics. Spatio-temporal knowledge graphs (STKGs) have therefore emerged as a powerful representation paradigm, as they integrate entities, relationships, time and space within a unified graph structure. They are increasingly applied across diverse domains, including environmental systems and urban, transportation, social and human mobility networks. However, modeling STKGs remains challenging: their foundations span classical graph theory as well as temporal and spatial graph models, which have evolved independently across different research communities and follow heterogeneous modeling assumptions and terminologies. As a result, existing approaches often lack conceptual alignment, generalizability and reusability.

This survey provides a systematic review of spatio-temporal knowledge graph models, tracing their origins in static, temporal and spatial graph modeling. We analyze existing approaches along key modeling dimensions, including edge semantics, temporal and spatial annotation strategies, temporal and spatial semantics and relate these choices to their respective application domains. Our analysis reveals that unified modeling frameworks are largely absent and that most current models are tailored to specific use cases rather than designed for reuse or long-term knowledge preservation. Based on these findings, we derive modeling guidelines and identify open challenges to guide future research.

\end{abstract}


\section{Introduction}

\subsection{Motivation}
\noindent

\setlength{\parskip}{0pt} 

\begin{figure}[b!]
    \centering
    \includegraphics[width=1\linewidth]{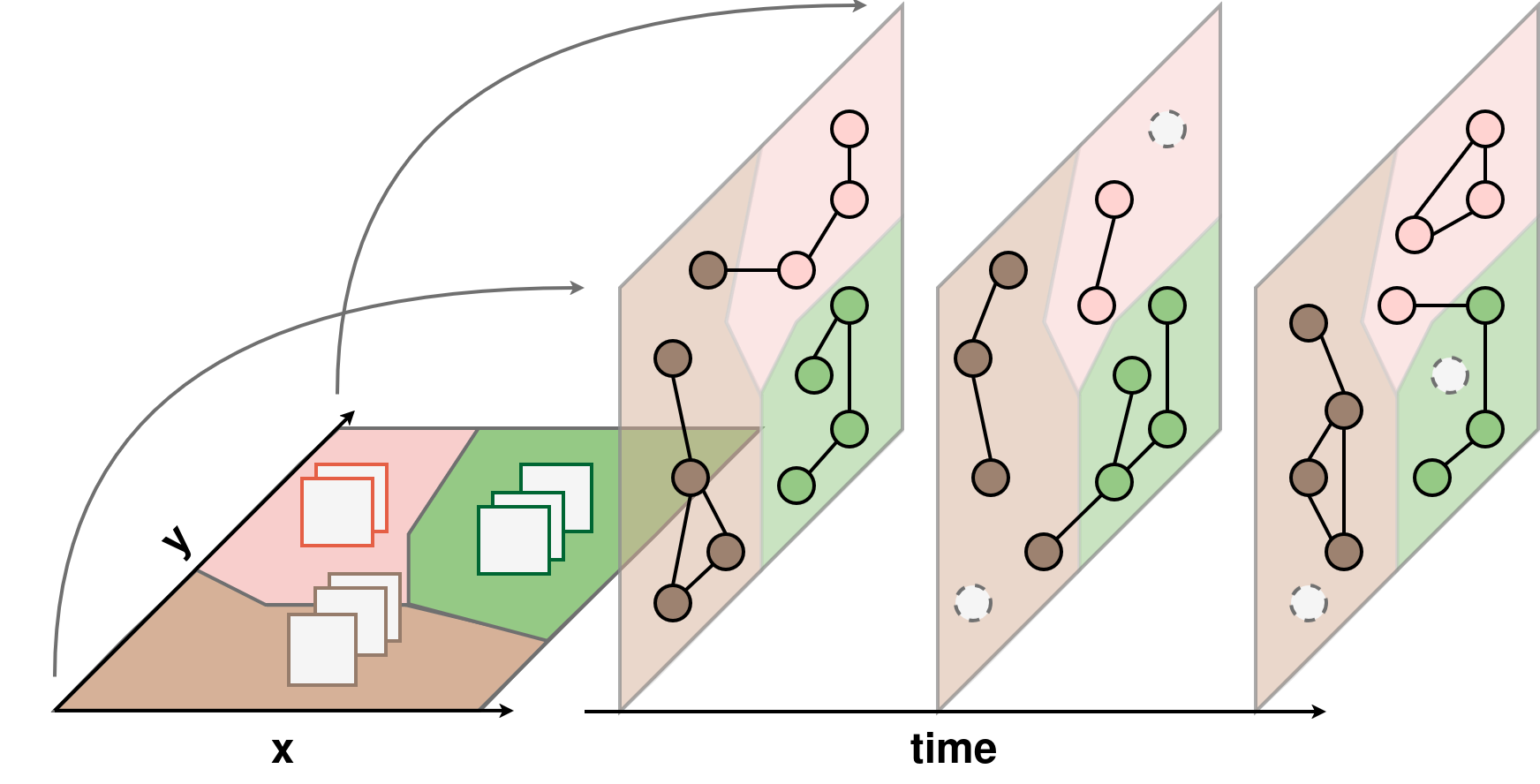}
    \caption{From spatio-temporal data to a spatio-temporal knowledge graph. In the example scenario, data is collected at multiple sites over several points in time (left). The spatio-temporal knowledge graph represents these data points as nodes, incorporating their spatial information, temporal attributes and possible relationships (right).}
    \label{fig:layer_approach}
\end{figure}

\noindent
Spatio-temporal knowledge graphs (STKGs) model complex real-world phenomena with inherent spatial and temporal structure. 
For instance, they are frequently applied in artificial intelligence for urban networks, where they provide a semantically rich framework to represent knowledge on a temporal and spatial scale \cite{Wang_2022, Jin_2024, Zeghina_2024}.
Figure \ref{fig:layer_approach} illustrates data collected at multiple sites over several points in time (left), which is transformed into an STKG representation (right). 
Temporal changes can affect the topology or properties at nodes and edges, as captured by the graph snapshots on the right. Further spatial properties allow entities to be located within space, e.g. in a cartesian coordinate system or a spatial reference system, affecting for instance the position and distance between nodes.
Graph structures play a key role in knowledge representation and are used in technologies such as deep learning \cite{Scarselli_2009, kipf_2016} and large language models \cite{Pan_2024}. While the combination of temporal and spatial properties can yield valuable insights and further expand the capabilities of these technologies, 
it also introduces significant challenges in knowledge graph modeling.

\IEEEpubidadjcol 

Several knowledge graph developments and applications concerning either temporal \cite{Blonder_2012, Casteigts_2012, Holme_2015, Holme_2012, Michail_2016, Wang_2019, Mo_2021, cai_2022, Rossetti_2018} or spatial \cite{Phillips_2015, Heckmann_2015, Cantwell_1993, Galpern_2011, Barth_lemy_2011} properties have been separately surveyed in the literature. Previous overviews of temporal and spatial graphs tend to emphasize their applications and interpretations rather than providing formal or standardized definitions of the underlying graph models.
This is notable given that the graph model determines the basic structure of a graph by fundamentally defining the meaning and properties of the nodes and edges. 
Despite the growing application of STKGs in diverse domains, existing approaches on temporal and spatial knowledge graph modeling are heterogeneous due to their origins and autonomous development. Since the start of graph theory, various research communities have independently developed analogous concepts and applied them across a wide range of fields, including the natural sciences, physics, geography and computer science. 
As a result, there are a variety of terminologies and models that have not been systematically reviewed or harmonized so far. Examining the origins of the diverse temporal and spatial graph models in depth could promote interdisciplinary exchange and support the development of standardized methods. For instance, automated knowledge graph generation methods could benefit from reusing and integrating STKG models into prompt engineering to ensure effective modeling of spatio-temporal dynamics. 
Based on consistent terminology and increased interoperability, uniform STKG models can facilitate the creation of long-term, mutually compatible knowledge graphs. 

\subsection{Application domains}

\noindent
To demonstrate the wide range of spatio-temporal graph applications, we highlight several relevant domains. In many fields, data is linked to spatial references, e.g. geographic coordinates and collected over time to monitor or predict trends and developments. When these data is also inherently interconnected, graph analysis methods can provide meaningful insights. Potential application areas include:

\begin{itemize}
    \item \textbf{Sensor networks} (e.g. urban, logistics or electricity networks): nodes represent spatially distributed sensors with associated time-series measurements and edges denote streets, transportation routes or transmission lines connecting them. Possible use cases include congestion prediction, route optimization and infrastructure planning.
    \item \textbf{Human interaction networks} (e.g. social or epidemiological networks): nodes represent individuals and their location at a specific time and edges denote relationships or proximity. Possible use cases include information spread modeling, recommendation and pattern matching.
    \item \textbf{Computer networks} (e.g. cybersecurity or router networks): nodes represent devices with their positions and current status and edges denote the movement of users or data packets between them. Possible use cases include network traffic control and user movement analysis.
    \item \textbf{Ecological networks} (e.g. environmental or biological networks): nodes represent measurements such as detected organisms associated to recording dates and regions, while edges represent  relationships or dependencies such as species interactions. Possible use cases include ecosystem stability and resilience analysis, biodiversity maintenance and nutrient cycling.  
\end{itemize}

\noindent
Recent research efforts have addressed a range of use cases for STKGs such as forest fire prediction \cite{Ge_2022}, maritime communication \cite{Zhang_2023}, traffic forecasting \cite{Guo_2019, Bui_2021}, urban mobility prediction \cite{Wang_2021}, mobile phone activity \cite{Li_2025}, leakage risk assessment in water distribution networks \cite{Wu_2024} and skeleton-based action recognition \cite{Yan_2018}. Throughout the survey additional applications are examined, emphasizing their significance across a variety of domains.

\subsection{Scope and contributions}

\noindent
Given the requirements and wide application of STKGs, our survey focuses on the modeling of spatio-temporal properties in knowledge graphs, with a particular focus on their evolution from both, temporal and spatial graph models. We present the foundational concepts of various graph modeling approaches and consolidate the multiple terminologies and models found in the temporal and spatial graph literature. 
Our goal is to organize this knowledge comprehensible across diverse research communities. Through this survey, we aim to address the broad scientific community of computer science, both as a field of research in its own and as an important interdisciplinary field supporting numerous applications across multiple domains.


This survey aims to consolidate the developments and concepts underlying spatio-temporal knowledge graph models. We here focus on the graph representation of temporal and spatial properties during modeling. 
Beyond the scope of this survey, practitioners need to decide during implementation which type of knowledge graph is best suited for a given use case, e.g. a property graph \cite{rodriguez_2010} or RDF graph \cite{manola_2004, hitzler_book_2008, Hitzler_2021} combined with ontologies \cite{Gruber_1993, Guarino_2009}. 

Existing surveys and approaches primarily focused on the application of spatio-temporal knowledge graphs in deep learning tasks \cite{Jin_2024, Zeghina_2024} or tried to formalize spatio-temporal data management specifically for RDF graphs \cite{Koubarakis_2010, Zhang_2021, Zhu_2020, Nayyeri_2022, Shbita_2023} or ontologies \cite{Hoffart_2013, Meng_2022, Zhu_2023}. 
To the best of our knowledge, this is the first survey dedicated to STKG models, where we trace the origins of STKGs in temporal and spatial knowledge graphs and review the current literature to harmonize and guide STKG modeling. The main contributions of this work are:

\begin{itemize}
    \item A systematic overview of spatio-temporal knowledge graphs and their development from static, temporal and spatial graph models across various application domains.
    \item An initial modeling guideline for integrating temporal and spatial properties into nodes and edges as basis for STKG development. 
    \item An overview of current and emerging challenges associated with spatio-temporal knowledge graphs.
\end{itemize}

\noindent
The remainder of this survey is structured as follows: Section \ref{sec:kg} offers an overview of static knowledge graphs as the foundation of subsequent models and research on temporal and spatial knowledge graph models. Section \ref{sec:stkg} reviews foundational and contemporary spatio-temporal knowledge graph models and a guideline for their modeling. Section \ref{sec:challenge} addresses current challenges encountered when working with spatio-temporal knowledge graphs and Section \ref{sec:conclusion} concludes the survey. 


\subsection{Survey methodology}

\noindent
The literature review covers a broad range of sources to provide a comprehensive overview including recent and  historically relevant publications. In preparing this survey, we aimed to identify the different types of graph models proposed in the literature. Therefore, we searched for studies on static, spatial, temporal and spatio-temporal graphs. In many cases, publications focus on graph analysis without explicitly discussing the underlying modeling approach, which required us to examine various papers in detail to determine whether a graph model was defined. We did not apply a strict time filter, as some graph models were introduced early in the research history and we want to ensure that no significant contributions are excluded.

We include  publications indexed in Google Scholar, Google, IEEE Xplore and Connected Papers. Our initial search terms are ``Knowledge Graph'', ``Temporal (Knowledge) Graph'', ``Temporal Network'', ``Spatial (Knowledge) Graph'', ``Spatial Network'', ``Spatio Temporal (Knowledge) Graph'' and ``Spatio Temporal Network''. We only include papers that explicitly propose a graph model to ensure that our survey covers foundational model frameworks rather than purely application-focused studies. We include publications written in English language. 
To ensure a comprehensive coverage of the relevant literature, we then applied a snowballing approach \cite{Wohlin_2014}. After identifying key publications, we examined their reference lists (backward snowballing) and reviewed papers that cited them (forward snowballing) to locate additional relevant studies.

\section{Knowledge graphs}\label{sec:kg}

\noindent
A ``graph'' generally refers to a set of entities (e.g. objects, persons or regions) that are connected through different types of relationships. The term ``knowledge graph'' does not yet have a universally accepted definition, as it is subject to multiple interpretations and ongoing debates regarding its scope and boundaries \cite{ehrlinger2016towards, Hogan_2021, Hofer_2024}. In this survey, we use the term ``knowledge graph'' in a broad sense to refer to a graph-based representation of entities and their relationships (such as a property graph or RDF graph) that may have additional metadata, e.g. time and space. We also use the terms ``graph'' and ``network'' interchangeably.   

\begin{figure}[h!]
    \centering
    \includegraphics[width=1\linewidth]{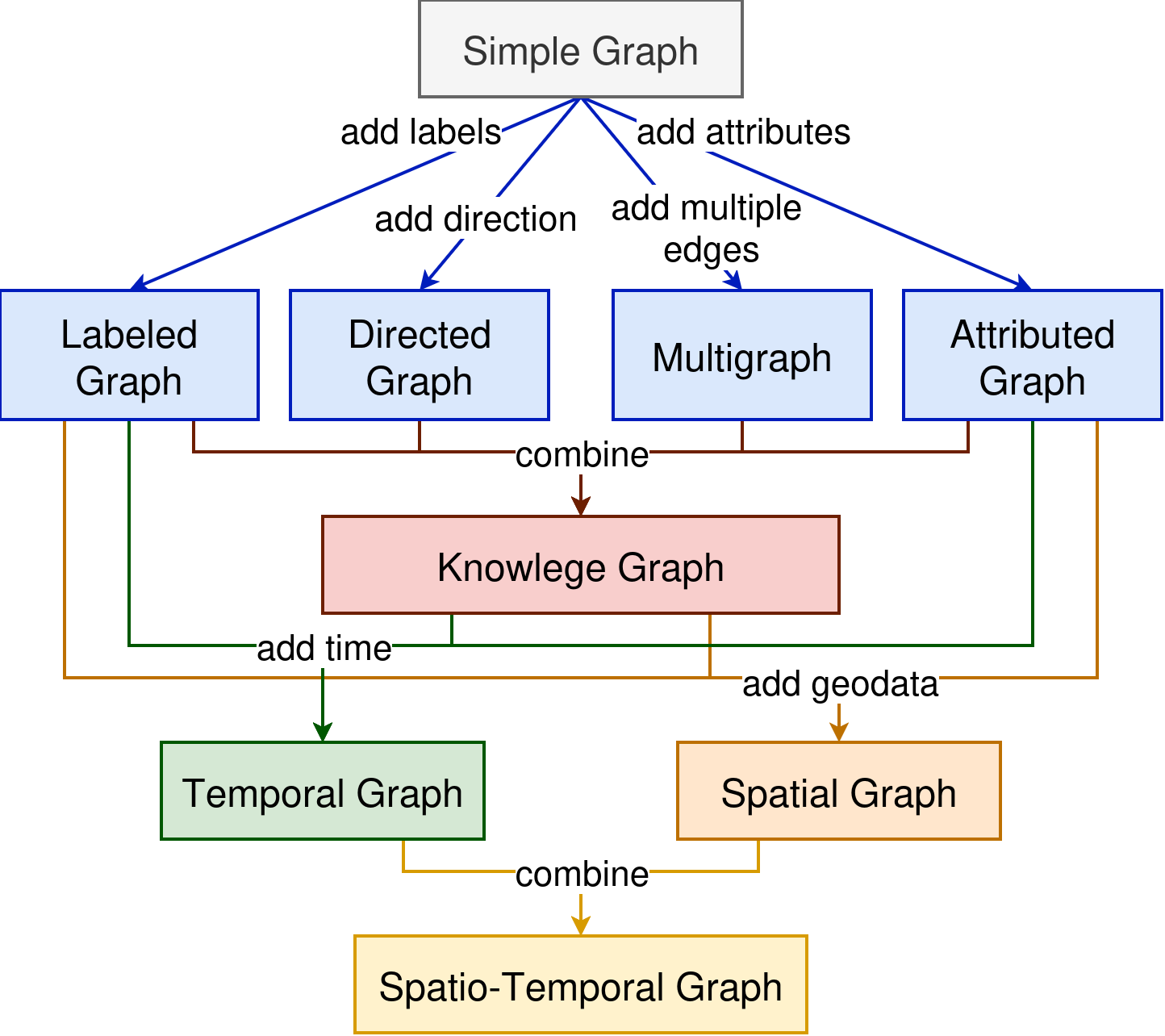}
    \caption{Overview of STKG origins. The simple graph model has been extended by additional characteristics and combines temporal and spatial properties.}
    \label{fig:simple_to_stkg}
\end{figure}

Figure \ref{fig:simple_to_stkg} provides a schematic overview of the development from simple graph models (top) to spatio-temporal graph models (bottom).
This section first introduces fundamental static graph models in Section \ref{sec:static_kg}, followed by approaches for incorporating temporal (Section \ref{sec:temporal_kg}) and spatial (Section \ref{sec:spatial_kg}) information.    

\subsection{Static graph models}\label{sec:static_kg}

\noindent
Graphs have been studied predominantly without semantically enriched nodes and edges, e.g. without temporal and spatial information. 
A wide range of application areas for such static graph models was for instance collected by Albert and Barabási \cite{Albert_2002}.
Definition \ref{def:simple_graph} presents a commonly used definition of a simple graph \cite{hackensack_1991, bondy_1976}, which remains the foundation for later advances in graph modeling.

\begin{definition}[Simple Graph]\label{def:simple_graph}
    A simple graph $G$ is described as $G = (V, E)$. $V$ is a nonempty set of vertices or nodes $v \in V$. $E$ is a possibly empty ($E = \emptyset$) unsorted set of undirected edges or relationships $E \subseteq \{\{u, v\} \mid u, v \in V, u \ne v\}$ connecting the vertices. 
\end{definition}

Some definitions of simple graphs further include an incidence function $\psi_G : E \rightarrow (V \times V)$ \cite{bondy_1976} which is a total function assigning a pair of vertices of $G$ to each edge of $G$. 
Graphs can be represented as incidence or adjacency matrices \cite{bondy_1976}: 
\begin{itemize}
    \item An incidence matrix $M(G)$ assigns to each vertex the edges it is incident to. It is a $V \times E$ matrix defined by $M(G)=[m_{ij}]$ where $m_{ij}$ is the number of times that $v_i \in V$ and $e_j \in E$ are incident.
    \item In an adjacency matrix  $A(G)$, edges are not represented explicitly but are implicitly determined by vertex pairs ($V \times V$). It is defined as $A(G) = [a_{ij}]$ where $a_{ij}$ is the number of edges connecting $v_i \in V$ and $v_j \in V$.
\end{itemize}


Although the simple graph model is widely used in many applications, it cannot capture information beyond the presence of nodes and edges between them. Therefore, simple graphs have been extended in several ways: 
\begin{itemize}
    \item In a directed graph edges are sorted pairs of nodes $(u,v)$.
    \item 
     In a multigraph two nodes can be interconnected by multiple edges and loops between them are allowed.
     \item Labeled graphs assign labels to nodes and edges \cite{dwyer_2016}. 
     \item An attributed graph adds attributes at the nodes or edges, e.g. as key-value pairs \cite{Wang_2014}. 
     \item A weighted graph covers intensities or weights of relationship between the nodes \cite{Barrat_2004, Newman_2004, hackensack_1991}, going beyond mere topology, i.e. whether an edge exists or not.
\end{itemize}

By incorporating those modifications, the property graph model and the RDF graph model extend the simple graph model to define semantically rich knowledge graphs. 
Both are widely used across various applications \cite{rodriguez_2010, hartig_2014, robinsons_2015, Junghanns_2016, Francis_2018} and are supported by a range of graph databases \cite{Besta_2023}. 
The following definitions \ref{def:property_graph} and \ref{def:rdf_graph} provide commonly accepted definitions of the property graph model \cite{angles2018property} and the RDF graph model \cite{Tamer_2016}: 

\begin{definition}[Property Graph \cite{angles2018property}]\label{def:property_graph}
    A property graph can be defined as $G = (N, E, \rho, \lambda, \sigma)$ where $N$ is a set of nodes, $E$ is a set of edges and $\rho$ describes the incidence function denoted as $\psi_G$ in the simple graph model. The partial functions $\lambda$ and $\sigma$ associate nodes and edges with labels and properties also referred to as keys and values. 
\end{definition}

\begin{definition}[RDF Graph \cite{Tamer_2016}]\label{def:rdf_graph}
    An RDF graph can be defined as $G=\langle V,L_V,f_V,E,L_E,f_E \rangle$ where $V$ is a set of vertices, $E$ is a set of edges, $L_V$ is a set of vertex labels, $L_E$ is a set of edge labels and $f_V$ ($f_E$) is an vertex (edge) labeling function that assign a label to each vertex (edge).
\end{definition}


\subsection{Temporal graph models}\label{sec:temporal_kg}

\noindent
Data collected in real-world systems is often dynamic rather than static including temporal information such as the time of recording or the last update. Temporal information allows to put data into context, e.g. to analyze developments and trends. Knowledge graphs that are created for temporal data management should therefore reflect the temporal information in the graph model. Temporal changes can affect for instance the topology of a graph, the properties at nodes and edges or the relationships between the nodes \cite{Holme_2015}. 
Several research efforts are directed toward integrating temporal extensions into graph databases \cite{Angles_2008, Angles_2012}, both in terms of temporal graph database models \cite{massri_2020, Hou_2024} and temporal graph query languages \cite{campos_2016, Debrouvier_2021}.  

In order to meet the complex requirements of temporal knowledge graph applications, existing graph analysis methods had to be adapted. For instance, there are temporal extensions for centrality measures \cite{Kim_2012} (such as the temporal ``PageRank'' algorithm \cite{Rozenshtein_2016}), for shortest path calculations \cite{Wu_2014, Huo_2014} and for detecting communities \cite{Speidel_2015, Rossetti_2018}. 

The rapid development of temporal knowledge graph models and analysis methods has led to a multitude of non-standardized names and terminologies, with authors repeatedly define similar or related concepts and models in different ways \cite{Holme_2012}. To provide an overview, we searched the literature on temporal knowledge graphs for closely related terms or synonyms and summarize them in Table \ref{tab:temporal_names}. 

\begin{table}[htbp]
\centering
\caption{Names of temporal knowledge graphs}
\label{tab:temporal_names}
\renewcommand{\arraystretch}{1.25} 
\begin{tabular}{|l|l|}
\hline
\textbf{Name}               & \textbf{Publication} \\ \hline
directed acyclic graph      & \cite{Speidel_2015} \\ \hline
dynamic networks            & \cite{Kuhn_2011} \\ \hline    
evolving graph              & \cite{Moffitt_2017, Ferreira_2004, Huo_2014} \\ \hline
evolving network            & \cite{impedovo_2017} \\ \hline
graphs over time            & \cite{Leskovec_2005} \\ \hline
interaction network         & \cite{Rozenshtein_2014, Kumar_2015, Rossetti_2015} \\ \hline
link stream graph           & \cite{Viard_2016, Latapy_2018} \\ \hline
temporal graph              & \cite{Wu_2014, Michail_2016, campos_2016, Gandhi_2020, Lou_2023, Semertzidis_2019, Steer_2020, Kostakos_2009, Khurana_2013} \\ \hline
temporal network            & \cite{Kempe_2002, Ahmed_2016, Holme_2012} \\ \hline
temporal property graph     & \cite{Debrouvier_2021, Byun_2020, Rost_2019} \\ \hline
time-dependent graph        & \cite{Wang_2019}  \\ \hline
time-evolving graph         & \cite{Sun_2007} \\ \hline
time-ordered graph          & \cite{Kim_2012}   \\ \hline  
time-varying graph          & \cite{basu_2010, Casteigts_2012, santoro_2011} \\ \hline
time-varying network        & \cite{Perra_2012} \\ \hline
\end{tabular}
\end{table}

The landscape of temporal knowledge graphs has expanded considerably and they are now widely used across many domains, as surveyed by Holme \cite{Holme_2015}.
Several adaptations of static graph models have been proposed for modeling temporal knowledge graphs. We surveyed the literature and identified different categories for 1) edge direction, 2) temporal annotation and 3) time semantics, which we outline in more detail in the following subsections. Later in this section, we use our categorization in Table \ref{tab:graph-models} to summarize temporal graph models according to those properties.  

\subsubsection{\textbf{Edge direction}}
In temporal knowledge graphs, edges can be undirected
, directed 
or a combination of both
. An undirected edge denotes an equal (symmetric) relationship between two nodes at a certain point in time (e.g. a connection between two computers in a computer network \cite{Ferreira_2004}), while a directed edge can represent a flow of information between nodes (e.g. in a citation network \cite{Rost_2019}).

\subsubsection{\textbf{Temporal annotation}} 
\begin{figure}[htbp]
    \centering
    \resizebox{1\linewidth}{!}{\begin{tikzpicture}[
roundnode/.style={circle, fill=node_inner, draw=node_outer, minimum size=10mm}
]
\node (a) at (0.5,4.25) {A)};  
\node (b) at (5,4.25) {B)};  
\node (c) at (0.5,2.5) {C)};  
\node (d) at (5,2.5) {D)};  
\node (GT) at (7,2.05) {$G_t$};  
\node[roundnode]        (a_left) at (1.5,4.25)       [] {$u_t$};
\node[roundnode]        (a_right)               [right=1cm of a_left] {$v_t$};
\node[roundnode]        (b_left)                [right=1.5cm of a_right] {$u$};
\node[roundnode]        (b_right)               [right=1cm of b_left] {$v$};
\node[roundnode]        (c_left)                [below=0.75cm of a_left] {$u_t$};
\node[roundnode]        (c_right)               [below=0.75cm of a_right] {$v_t$};
\node[roundnode]        (d_left)                [below=0.75cm of b_left] {$u$};
\node[roundnode]        (d_right)               [below=0.75cm of b_right] {$v$};

\draw[-] (a_left.east) -- (a_right.west);
\draw[-] (b_left.east) -- node[anchor=south] {$e_t$} (b_right.west);
\draw (0,3.5) -- (9,3.5);
\draw (4.5,1.5) -- (4.5,5);
\draw[-] (c_left.east) -- node[anchor=south] {$e_t$} (c_right.west);
\draw[-] (d_left.east) -- (d_right.west);
\draw (5.35,3.15) -- (8.75,3.15) -- (8.75,1.85) -- (5.35,1.85) -- cycle;
\end{tikzpicture}}
    \caption{Temporal annotations in a temporal knowledge graph. In a node-annotated graph (A) the time is stored at the nodes, i.e. the nodes $u$ and $v$ are valid at time $t$. 
    An edge-annotated graph (B) stores the time at the edges, i.e. the edge between $u$ and $v$ exists at time $t$, the nodes remain static. 
    In a node-edge-annotated graph (C) the time can be stored at the nodes and edges. 
    In a graph-annotated approach (D) neither the nodes nor the edges store time. The graph $G_t$ represents the nodes and edges at time $t$.}
    \label{fig:annotations_temp}
\end{figure}
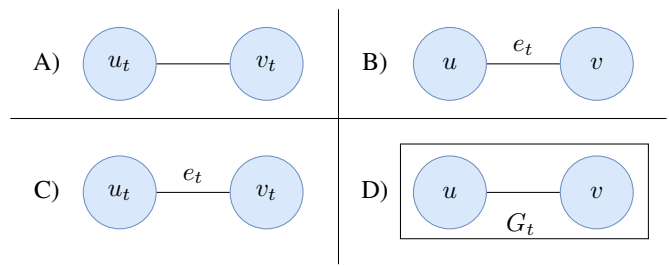

Any temporal knowledge graph model must provide a way to represent temporal information. For instance the temporal extension of the property graph model (see Definition \ref{def:property_graph}) is often referred to as a temporal property graph \cite{Byun_2020, Debrouvier_2021, Andriamampianina_2022}. Overall, we identified four main possibilities in the literature to incorporate time in a temporal knowledge graph, as illustrated in Figure \ref{fig:annotations_temp}: A) node-annotated, 
B) edge-annotated, 
C) node-edge-annotated 
and D) graph-annotated. 

Campos et al. \cite{campos_2016} propose a \textbf{node-annotated} temporal graph model, i.e. only the nodes store temporal information. The authors demonstrate its applicability for social networks. They define a temporal graph as $G(N_o, N_e, N_a, N_v, E)$, where $E$ is a set of edges and $N_o, N_e, N_a$ and $N_v$ are denoted as object, edge, attribute and value nodes. Each node is assigned a tuple $(name, interval)$ that contains the name of the node and the interval in which the node is valid. 
The model therefore has predefined types of nodes that store different attributes. 
The edges don't store any information instead they connect the different types of nodes following defined constraints. 

A purely \textbf{edge-annotated} temporal graph model was proposed for social networks by Rossetti et al. \cite{Rossetti_2015}. 
In their temporal graph $G = (V, E, T)$, temporal properties are added solely to edges. $V$ is a set of static vertices, $T$ is a set of timestamps and $E$ is a set of timestamped edges. Each edge $e \in E$ is described by a tuple $(u,v,t)$ with its starting ($u \in V$) and ending ($v \in V$) vertex and the timetamp $t \in T$ when the edge is valid. 

To allow for the  enrichment with temporal information both at nodes and edges, Moffitt and Stoyanovich \cite{Moffitt_2017} proposed a \textbf{node-edge-annotated} temporal graph model that is build on top of the property graph model by Angles \cite{angles2018property} (see Definition \ref{def:property_graph}). The authors define a temporal graph as $G = (V, E, L, \rho, \xi^T, \lambda^T)$, where $V$ is a finite set of vertices, $E$ is a finite set of edges, $L$ describes a finite set of property labels and $\rho$ is the incidence function. Additionally the total function $\xi^T$ maps a node (edge) together with a time period to a boolean to indicate the existence of the node (edge) during the time period. Also the partial function $\lambda^T$ maps a node (edge) together with a property label and a time period to a value of the property during the time period.
The authors demonstrated the applicability of the model for interaction and further networks.

Finally, in a \textbf{graph-annotated} temporal graph model, as proposed by Semertzidis and Pitoura \cite{Semertzidis_2019}, 
a temporal graph $G_{[t_i,t_j]}$ is a sequence of graph snapshots $\{G_{t_i}, G_{t_i+1}, \ldots, G_{t_j}\}$. Each  graph snapshot $G_t = (V_t, E_t, L_t)$ represents the active set of vertices  $V$, edges $E$ and labels $L$ at a discrete timestamp $t$. 
This temporal graph annotation was adopted, for example, to biological networks \cite{Semertzidis_2019}.

\newcommand*\rot{\rotatebox{90}}
\begin{table*}[t]
\centering
\caption{Temporal knowledge graph models}
\vspace{-3mm}
\footnotesize{Focus: D - Application Domain, M - Generic Graph Model, A - Analysis/Algorithm}  
\bigskip

\label{tab:graph-models}
\setlength{\tabcolsep}{0.6em}
\renewcommand{\arraystretch}{1.25} 
\begin{tabular}{l|l|cc|ccc|ccc|c|l}
\multicolumn{1}{c|}{\textbf{}} & \multicolumn{1}{c|}{\textbf{}} & \multicolumn{2}{c|}{\textbf{Edges}} & \multicolumn{3}{c|}{\textbf{\begin{tabular}[c]{@{}c@{}}Temporal \\ Annotation\end{tabular}}} & \multicolumn{3}{c|}{\textbf{\begin{tabular}[c]{@{}c@{}}Temporal \\ Semantic\end{tabular}}} & \multicolumn{1}{c|}{\textbf{}} & \multicolumn{1}{c}{\textbf{}} \\
\multicolumn{1}{c|}{\textbf{Paper}} & \multicolumn{1}{c|}{\textbf{Year}} & \multicolumn{1}{l}{\rot{Undirected}} & \multicolumn{1}{l|}{\rot{Directed}} & \multicolumn{1}{l}{\rot{Node}} & \multicolumn{1}{l}{\rot{Edge}} & \multicolumn{1}{l|}{\rot{Graph}} & \multicolumn{1}{l}{\rot{Duration}} & \multicolumn{1}{l}{\rot{Interval}} & \multicolumn{1}{l|}{\rot{Timestamp}} & \multicolumn{1}{c|}{\textbf{Focus}} & \multicolumn{1}{c}{\textbf{Application Domain}} \\ \hline
Ferreira et al. \cite{Ferreira_2004} & 2004 & x & & & & x & & & x & M & wireless communication networks \\
Basu et al. \cite{basu_2010} & 2010 & x & x & & & x & & & x & A & dynamic random graphs \\
Kim et al. \cite{Kim_2012} & 2012 & x & x & & x & & & x & & A & human contact networks \\
Huo et al. \cite{Huo_2014} & 2014 & x & x & x & x & & & x & & A & social networks \\
Wu et al. \cite{Wu_2014} & 2014 & x & x & & x & & x & & x & A & flight graphs and social, bibliographic networks \\
Kumar et al. \cite{Kumar_2015} & 2015 & x & & & x & & & & x & A & social, bibliographic networks \\
Rossetti et al. \cite{Rossetti_2015} & 2015 & x & & & x & & & & x & A & social, bibliographic networks \\
Campos et al. \cite{campos_2016} & 2016 & x & & x & & & & x & & M & social networks \\
Moffitt and Stoyanovich \cite{Moffitt_2017} & 2017 & & x & x & x & & & & x & M & interaction, bibliographic networks \\
Semertzidis and Pitoura \cite{Semertzidis_2019} & 2019 & & x & & & x & & & x & A & bibliographic, biological networks \\
Rost et al. \cite{Rost_2019} & 2019 & & x & x & x & x & & x & & M & bibliographic, (synthetic) social networks \\
Gandhi et al.\cite{Gandhi_2020} & 2020 & & x & x & x & & & x & & M & social, bibliographic networks \\
Byun et al. \cite{Byun_2020} & 2020 & & x & x & x & & & x & x & M & bibliographic, computer, interaction networks \\
Debrouvier et al. \cite{Debrouvier_2021} & 2021 & & x & x & x & & & x & & M & social networks, flight graphs \\
Lou et al. \cite{Lou_2023} & 2023 & x & & & x & & & & x & A & bibliographic, money transfer networks \\
Plamper et al. \cite{plamper2023snapshot} & 2023 & & x & x & & & & & x & M & chemical networks \\
\end{tabular}
\end{table*}

\subsubsection{\textbf{Time semantics}}
\begin{figure}[b!]
    \centering
    \resizebox{\linewidth}{!}{\begin{tikzpicture}[
roundnode/.style={circle, fill=node_inner, draw=node_outer, minimum size=10mm}
]
\node[roundnode]        (a1) at (2,5.5)    [] {$a$};
\node[roundnode]        (a2) at (1,4)      [] {$b$};
\node[roundnode]        (a3) at (3,4)      [] {$c$};
\draw[->] (a1.west) .. controls (1,5.25) .. (a2.north);
\draw[->] (a1.east) .. controls (3,5.25) .. (a3.north);
\draw[->] (a2.east) -- (a3.west);
\node (a1) at (0.75,5) {3};  
\node (a2) at (3.25,5) {5};  
\node (a3) at (2,3.5) {1}; 

\node[roundnode]        (b1) at (5.75,5.5)    [] {$a$};
\node[roundnode]        (b2) at (4.75,4)      [] {$b$};
\node[roundnode]        (b3) at (6.75,4)      [] {$c$};
\draw[->] (b1.west) .. controls (4.75,5.25) .. (b2.north);
\draw[->] (b1.east) .. controls (6.75,5.25) .. (b3.north);
\draw[->] (b2.east) -- (b3.west);
\node (b1) at (4.5,5.65) {[73-00]};  
\node (b2) at (6.75,3.25) {[80-Now]};  
\node (b5) at (7,5.65) {[00-21]};
\node (b6) at (5.75,4.75) {[60-Now]};
\node (b7) at (4.75,3.25) {[63-Now]};

\node[roundnode]        (c1) at (9.55,5.5)    [] {$a$};
\node[roundnode]        (c2) at (8.55,4)      [] {$b$};
\node[roundnode]        (c3) at (10.55,4)      [] {$c$};
\draw[->] (c1.west) .. controls (8.55,5.25) .. (c2.north);
\draw[->] (c1.east) .. controls (10.55,5.25) .. (c3.north);
\draw[->] (c2.east) -- (c3.west);
\node (a1) at (8.2,5) {1,2};  
\node (a2) at (10.8,5) {2};  
\node (a3) at (9.55,3.5) {1,3};  

\draw (3.75,3) -- (3.75,7);
\draw (7.75,3) -- (7.75,7);
\node (t1) at (2,6.5) {A) Duration};  
\node (t2) at (5.75,6.5) {B) Interval};  
\node (t3) at (9.55,6.5) {C) Timestamp};
\end{tikzpicture}}
    \caption{Temporal semantics in a temporal knowledge graph. A duration-labeled semantic (A) defines the traversal time from one node to another. An interval-labeled semantic (B) indicates the periods during which the nodes or edges are valid. A timestamp-labeled semantic (C) reflects discrete timestamps at which a node or edge is valid.}
    \label{fig:labeled}
\end{figure}
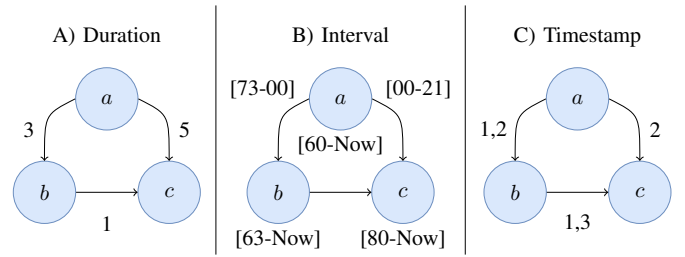

Time semantics describes how time is to be interpreted in a temporal knowledge graph.
Figure \ref{fig:labeled} illustrates different time semantics used in existing temporal knowledge graph models: time can refer to A) a duration, 
B) an interval  
or C) a timestamp.

Wu et al. \cite{Wu_2014} proposed a \textbf{duration-labeled} time semantic (Figure \ref{fig:labeled} A). The authors extend a simple graph model with directed temporal edges $e=(u,v,t,\lambda)$ where $t$ is the starting time and $\lambda$ is the traversal time from $u$ to $v$. This time semantic can for instance be used to represent flight durations in a flight graph or phone call durations in a phone call graph.

An \textbf{interval-labeled} time semantic was proposed by Debrouvier et al. \cite{Debrouvier_2021} (Figure \ref{fig:labeled} B). 
They define a temporal knowledge graph, based on Campos et al. \cite{campos_2016}, as $G(N_o, N_a, N_v, E)$. In the model the edges $E$, the object nodes $N_o$, the attribute nodes $N_a$ and the value nodes $N_v$ are associated with intervals representing their respective valid time periods. Defined constraints determine which types of nodes may be connected to each other. 
A special value $now$ indicates whether a node or edge is currently valid \cite{Clifford_1997}. 
This time semantic can, for instance, be applied to temporal pattern matching in a bicycle rental graph \cite{Rost_2021}.

A \textbf{timestamp-labeled} time semantic was proposed by Lou et al. \cite{Lou_2023}. The authors extend the edges of a simple graph by representing each edge $e \in E$ as a triplet $e=(u,v,t)$, where $u$ and $v$ are static nodes and $t$ is the timestamp at which the edge is active. This time semantic enables the graph to be represented as a sequence of static graphs, often referred to as snapshot graphs \cite{Semertzidis_2019, Huo_2014, Lou_2023, Ferreira_2004, basu_2010, Rossetti_2015, Kumar_2015}. The sequence of static graphs is well-suited for certain traditional analysis methods, which often rely on static graphs, making this approach a popular choice for graph modeling \cite{Dhouioui_2014, Sun_2007, Zhuang_2019}. A timestamp-labeled time semantic can be applied, for instance, in money transfer networks \cite{Sun_2022} or bibliographic networks \cite{Lou_2023}.

\subsubsection{\textbf{Summary}}
Modeling temporal knowledge graphs primarily involves three key aspects: (1) the edge direction, (2) the place to annotate the time and (3) the time semantic. Table \ref{tab:graph-models} summarizes how temporal knowledge graphs have been previously modeled. For each publication (row), the table indicates the year, whether edges are directed or undirected, where time is annotated, the semantics of time, the focus and application domains. The focus of a publication may lie in addressing a research question within a specific application domain (D), proposing a generic graph model (M) or introducing new generic analysis methods or algorithms (A). Even if a concrete focus is not always clearly distinguishable and may overlap, the categorization captures more general tendencies and provides a rough overview of the scope of each publication. 
For generic approaches that do not focus on specific use cases, the mentioned application domains in the right column of Table \ref{tab:graph-models} rather denote the domain(s) of evaluation data sets or running examples within the publication. According to the table, directed edges are the most commonly used in temporal knowledge graph models, followed by undirected edges, while some models support both types. In most cases, authors choose to annotate temporal information on both nodes and edges or solely on edges. Assigning temporal information exclusively to nodes or at a graph-level is relatively uncommon. Time is predominantly represented using discrete timestamps and less frequently as intervals. Duration-labeled time semantics are rare and typically appear only in combination with timestamp-labeled time semantics. The publications place equal emphasis on generic temporal graph models (M) and new analysis methods (A). In contrast, no application-oriented contributions (D) were identified. Some application domains appear frequently in the literature, such as social and bibliographic networks, whereas others occur only sporadically, including flight graphs, biological networks and chemical networks.

\subsection{Spatial graph models}\label{sec:spatial_kg}

\begin{table}[b]
\centering
\caption{Names of spatial knowledge graphs}
\label{tab:spatial_names}
\renewcommand{\arraystretch}{1.25} 
\begin{tabular}{|l|l|}
\hline
\textbf{Name}               & \textbf{Publication} \\ \hline
geospatial network          & \cite{Wang_2020} \\ \hline
geo-social network          & \cite{scellato_2010, Wang_2016, Hristova_2016, Fang_2017} \\ \hline
landscape graph             & \cite{Cantwell_1993} \\ \hline
patch-based graph           & \cite{Galpern_2011} \\ \hline
power grid                  & \cite{Pagani_2013} \\ \hline
spatial network             & \cite{Barth_lemy_2011, Zhong_2014} \\ \hline
spatial graph               & \cite{Fall_2007} \\ \hline
street network              & \cite{Jiang_2004, Buhl_2006} \\ \hline
traffic network             & \cite{Ding_2019} \\ \hline
transportation graph        & \cite{Kurant_2006_1} \\ \hline
\end{tabular}
\end{table}


\noindent
In real-world systems spatial information such as geographic coordinates play a crucial role, since  many processes, interactions, and dependencies are influenced by the positions of entities in space. 
For instance, spatial information enables the visualization of data on a map or the analysis of distances between objects. Knowledge graphs can support spatial data management by modeling spatial relationships between entities such that their positioning is no longer arbitrary within an abstract space \cite{Costa_2007}.  

Fall et al. \cite{Fall_2007} define a spatial graph as a geometric reference system that links nodes and edges to specific spatial locations. This is one of the earliest modern references to spatial graph theory. In other domains, the term ``spatial graph'' has different meanings, which we do not examine in detail here. For instance, in mathematics it denotes a graph embedded in three-dimensional space, where edges may intersect \cite{Conway_1983, Negami_1991}.

The origins of spatial graphs date back to the field of quantitative geography in the mid-20th century \cite{Mackay_1959, Murray_2010, Barth_lemy_2011, Barthelemy_2018, Barthelemy_2022}, where researchers investigated models and methods for integrating mathematical graph theory into their field of research \cite{ROBINSON_1957, hagget_1969, Golledge_1978}.
Quantitative geography encompasses all methods and techniques used to study spatial phenomena, issues and problems, e.g. remote sensing, mathematics and geographic information systems (GIS) \cite{Murray_2010}. 
GIS are systems that specialize in the collection, storage, processing and presentation of spatial data \cite{fischer_2006} and serve as a basic framework to analyze spatial data \cite{longley_2015, bartelme_2005, Yano_2000}.
Although spatial graphs originate in geography, some scholars view them as abstract models, and refer to ``geographical graphs'' to emphasize place and meaning \cite{Uitermark_2021}.

\begin{figure}[t!]
    \centering
    \includegraphics[width=0.85\linewidth]{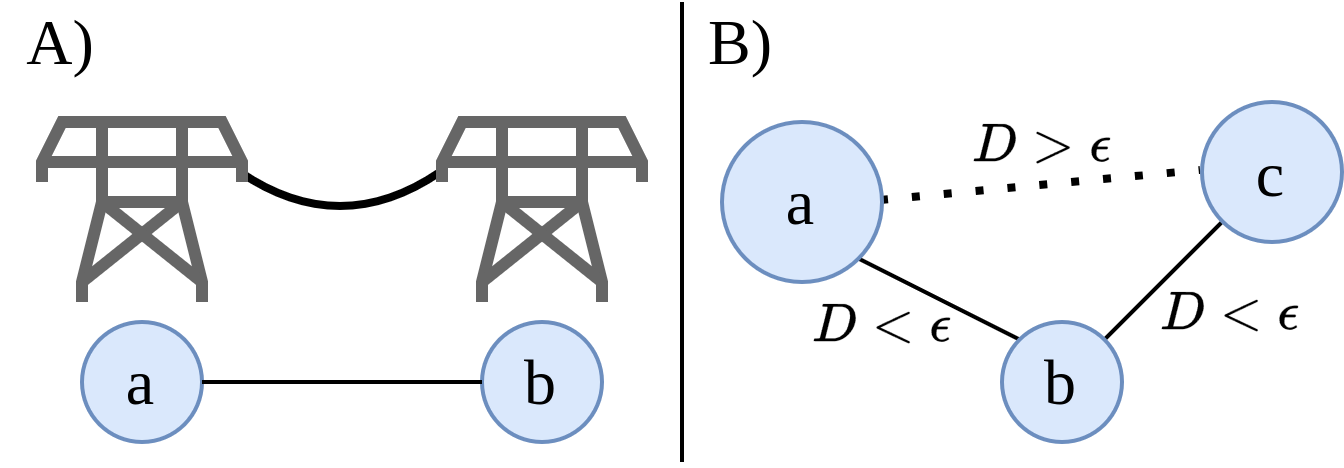}
    \caption{Obtaining edges between nodes with spatial information. Sometimes (A) edges are inherently given between nodes, e.g. as power lines between power poles. Inference methods (B) try to infer missing or unknown edges between entities, e.g. by defining a threshold for a distance.}
    \label{fig:edge_construction}
\end{figure}

Spatial graphs have a wide range of applications, including social networks \cite{Hristova_2016, Fang_2017, scellato_2010, Wang_2016}, transportation \cite{De_Montis_2007, Cardillo_2006, Buhl_2006, Kurant_2006_1, Ding_2019}, sports \cite{Buld__2018, Buld__2019}, public facilities \cite{Gastner_2006}, ecology \cite{Runions_2005, Shanthala_Devi_2013, Galpern_2011, Cantwell_1993}, energy \cite{Pagani_2013} and chemistry \cite{Danel_2020}. 
They are rarely called ``spatial graph'' and are usually named according to their specific applications as summarized in Table \ref{tab:spatial_names}, complicating literature search and selection. 
We categorize the following approaches to incorporate spatial information in knowledge graphs: 1) edge direction and inference, 2) spatial annotation and 3) dimensionality.
We first discuss the main approaches and then categorize the publications, as summarized in Table \ref{tab:graph-models-spatial}.

\subsubsection{\textbf{Edge direction \& inference}}
Spatial knowledge graphs are modeled using directed edges
, undirected edges 
or both type of edges
. Directed edges can model paths between entities under asymmetric weights, e.g. least-cost paths that account for landscape slope \cite{Fall_2007}. Undirected edges can represent spatial proximity, such as the distance between users in social networks \cite{Hristova_2016}. Sometimes edges can represent relationships that exist as physical entities, e.g. power lines in electrical grid networks (Figure \ref{fig:edge_construction} A) \cite{Pagani_2013}. The adjacency matrix of such topology-based graphs \cite{Jin_2024} can be generated if two nodes $v_i$ and $v_j$ are connected by a physical entity: 

\begin{equation}
   a_{ij} = \begin{cases} 
       1, & \text{if } v_i \text{ connects to } v_j, \\ 
       0, & \text{otherwise}. 
   \end{cases}
   \label{eq:adjacency_matrix}
\end{equation}

\noindent
In certain spatial knowledge graphs the topology is either ambiguous or absent. Edges can be missing or unknown despite the presence of nodes with spatial information, for instance in landscape modeling \cite{Urban_2001, Calabrese_2004} or internetworking \cite{Zegura_1996}. Several methods have been proposed to infer edges based on certain assumptions or rules. A detailed overview of these methods is provided by Barthélemy \cite{Barth_lemy_2011}. For instance the distance-based graph method \cite{Jin_2024, Chai_2018} is based on the Waxman model \cite{Waxman_1988} and, symbolically, on the first law of geography by Tobler \cite{Tobler_1970}, i.e. ``everything is related to everything else, but near things are more related than distant things''. The adjacency matrix of a distance-based graph can be generated if the distance $D$, e.g. euclidean, between two nodes $v_i$ and $v_j$ is smaller than a threshold value $\epsilon$ (Figure \ref{fig:edge_construction} B): 

\begin{equation}
   a_{ij} = \begin{cases} 
       1, & \text{if } D(v_i, v_j) < \epsilon, \\ 
       0, & \text{otherwise}. 
   \end{cases}
   \label{eq:distance_rule}
\end{equation}
\medskip

\begin{figure}[b]
    \centering
    \resizebox{0.6\linewidth}{!}{\begin{tikzpicture}[
roundnode/.style={circle, fill=node_inner, draw=node_outer, minimum size=12mm}
]
\node (a) at (0.5,4) {A)}; 
\node (c) at (0.5,2.5) {B)};
\node (b) at (0.5,0.75) {C)};   
\node (GT) at (3.15,0.22) {$G_{(x,y)}$};  
\node[roundnode]        (a_left) at (1.5,4)       [] {$u_{(x,y)}$};
\node[roundnode]        (a_right)               [right=2cm of a_left] {$v_{(x,y)}$};
\node[roundnode]        (c_left)                [below=0.3cm of a_left] {$u_{(x,y)}$};
\node[roundnode]        (c_right)               [below=0.3cm of a_right] {$v_{(x,y)}$};
\node[roundnode]        (d_left)                [below=0.5cm of c_left] {$u$};
\node[roundnode]        (d_right)               [below=0.5cm of c_right] {$v$};

\draw[-] (a_left.east) -- (a_right.west);
\draw (0,3.25) -- (5.5,3.25);
\draw (0,1.75) -- (5.5,1.75);
\draw[-] (c_left.east) -- node[anchor=south] {$e_{(p_1,p_2, \ldots,p_k)}$} (c_right.west);
\draw[-] (d_left.east) -- (d_right.west);
\draw (0.8,1.5) -- (5.4,1.5) -- (5.4,0) -- (0.8,0) -- cycle;
\end{tikzpicture}}
    \caption{Spatial annotations in a spatial knowledge graph. In a node-annotated graph (A) the spatial information is stored at the nodes, e.g. $u_{(x,y)}$. In a node-edge-annotated graph (B) the spatial information is additionally stored at the edges. Each point $p$ in $(p_1,p_2, \ldots,p_k)$ represents a coordinate $(x,y)$ and influences the path of the edge. In a graph-annotated approach (C) a group of nodes form a non-spatial graph but the group itself is annotated with spatial information.}
    \label{fig:annotations_spatial}
\end{figure}
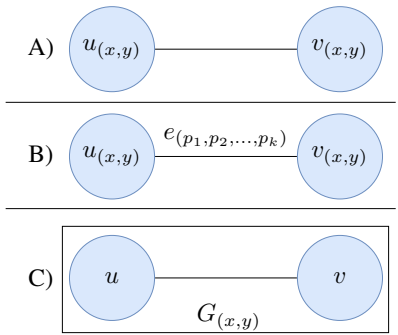

\noindent

\subsubsection{\textbf{Spatial annotation}}
Similar to temporal graphs, also spatial information can be modeled in different ways: A) node-annotated, 
B) node-edge-annotated and 
C) 
graph-annotated. 
Figure \ref{fig:annotations_spatial} illustrates the different modeling options for spatial knowledge graphs. 

\begin{figure}[t]
    \centering
    \includegraphics[width=0.65\linewidth]{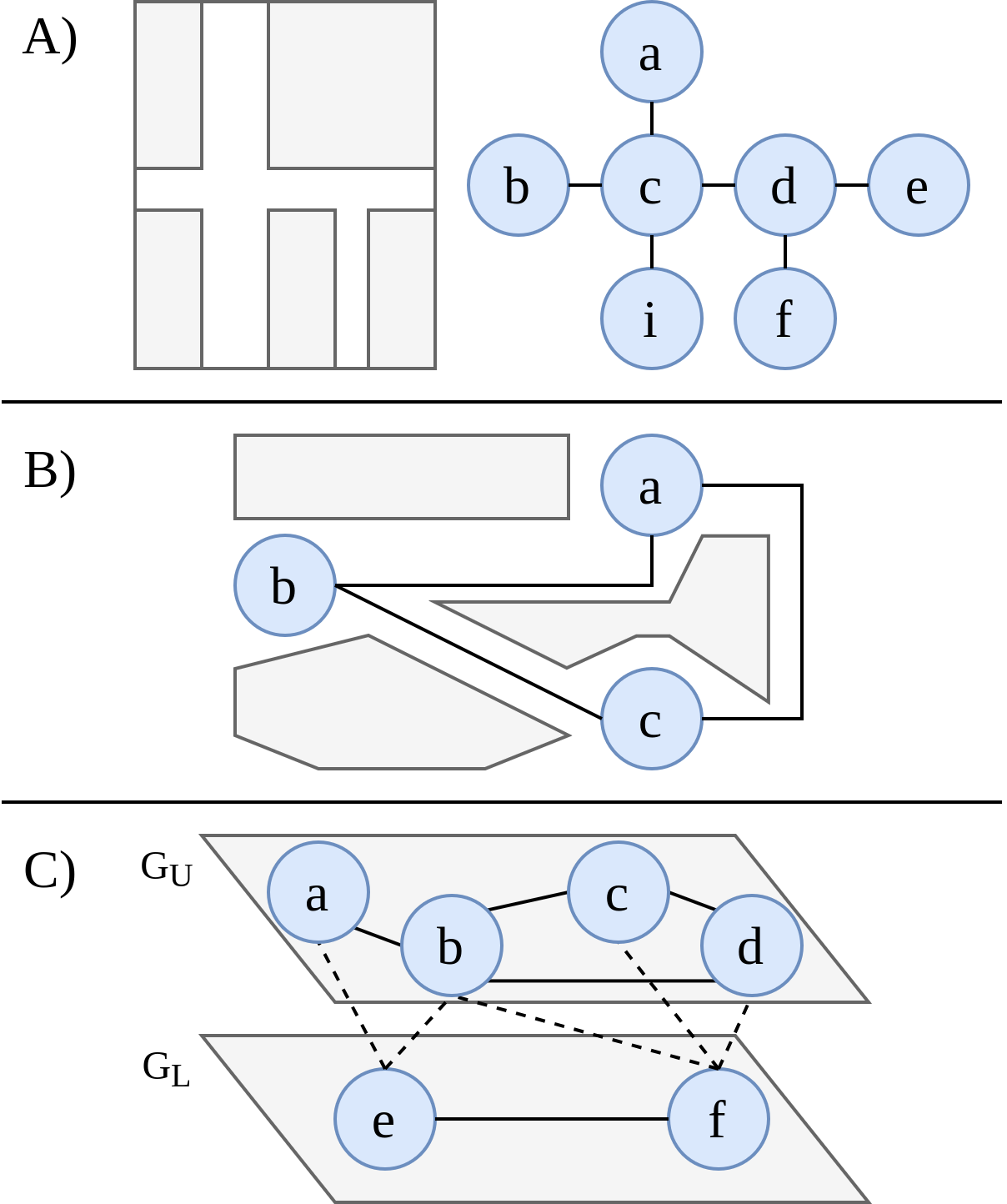}
    \caption{Examples of spatial knowledge graph models. A street graph (A) can be represented as a binary map (left) with public space (white) and private space (grey). The derived graph (right) considers intersections and dead-ends as nodes with coordinates and the street as edges between them. In a landscape graph (B) nodes can represent regions and edges the least-cost paths. A social network (C) can be divided into an upper layer $G_{U}$ that shows users and their relationships and a lower layer $G_L$ that shows locations and movement of the users from the upper layer. The edges between layers indicate where users have checked in.}
    \label{fig:spatial_examples}
\end{figure}

Buhl et al. \cite{Buhl_2006} proposed a \textbf{node-annotated} spatial graph model for street networks. The model extends earlier approaches, that omitted spatial information \cite{Jiang_2004, Rosvall_2005}, to enable the analysis of both topological and geometric patterns. The authors define an embedded planar graph $G =(V, E)$, where $V=\{(v_i,x_i,y_i), (i=1, \ldots, n)\}$ is a set of nodes characterized by position $(x,y)$ and $E = \{(v_i,v_j)\}$ is a set of edges between nodes. A planar graph is a graph that can be drawn on the plane without edges crossing each other \cite{nishizeki_1988}. Figure \ref{fig:spatial_examples} A illustrates a planar graph derived from a street network. The nodes are squares, street intersections and dead ends. The edges are sub-sections of streets.  

A \textbf{node-edge-annotated} spatial graph model was proposed by Fall et al. \cite{Fall_2007}. 
The authors define a spatial graph as $G = (N, L, S, f_{cost})$ where $N$ is a set of nodes, $L$ is a set of links (edges), $S$ is a spatial domain and $f_{cost}$ is a spatial cost function.  
A node $n$ is a two-dimensional area (region) consisting of contiguous square cells within a rectangular grid, referred to as spatial domain $S$, on a Cartesian plane. 
A link $l = ((n, m), (p_n,...,p_m))$ consists of the nodes $n$ and $m$ and a line between points $p_n$ and $p_m$. The points $p_n$ and $p_m$ are dimensionless locations inside the nodes $n$ and $m$.
A simple line is a one-dimensional vector between two points $(p_1, p_2)$. A general line is defined as a sequence of simple lines, denoted by a list of points $(p_1, p_2, ..., p_k)$ and influences the path between nodes. 
For instance, in ecology these paths can represent landscape connectivity \cite{taylor_1993}. 
A spatial cost function $f_{cost}$ is a spatially varying function over every square cell. The value is interpreted as the cost of traversal to reflect for instance movement energy or mortality risk. 
Figure \ref{fig:spatial_examples} B illustrates such a node-edge-annotated graph. The nodes represent places and the edges movement paths between them. The direct route is not always possible due to obstacles, that can be modeled as higher costs using the cost function $f_{cost}$. 

A \textbf{graph-annotated} spatial graph model was proposed by Hristova et al. \cite{Hristova_2016}. The authors describe an approach to project a social graph onto a spatial graph and vice versa. The model consists of two layers. The first layer is the social graph $G_U = (V_U, E_U)$, where $V_U$ is a set of users and $E_U$ is a set of friendship relations. The second layer is the spatial graph $G_L = (V_L, E_L)$, where $V_L$ is a set of locations and $E_L$ is a set of edges that describe user transitions between the locations.
An interconnected graph $G_M = (G_U, G_L, I)$ combines the social and the spatial graph, where $I$ is a set of edges that store the number of visits $w_{u,l}$ of a user $u \in V_U$ at a location $l \in V_L$. Figure \ref{fig:spatial_examples} C illustrates a graph-annotated spatial graph. The upper layer graph $G_U$ shows users and their relationships without spatial information. The lower graph $G_L$ represents locations with spatial information and movements of users from the upper graph. The edges between the layers depict places that users visit. 

\subsubsection{\textbf{Dimensionality}}
\begin{figure}[b]
    \centering
    \resizebox{0.9\linewidth}{!}{\begin{tikzpicture}[
roundnode/.style={circle, fill=black, draw=black, minimum size=1mm}
]
\draw (1.25,1.5) -- (2,1.5);
\draw (1,0.5) -- (2.5,0.5);
\draw[fill=node_inner, draw=node_outer] plot[smooth cycle] coordinates {(0, 0) (1, 0.5) (1.5, 1) (1, 2) (0, 1.5)};
\draw[fill=node_inner, draw=node_outer] plot[smooth cycle] coordinates {(2.5, 0.25) (3, 0.5) (3.5, 0.75) (3, 1.5) (2, 1.5)};
\node (a) at (1.75,1.65) {a};
\node (b) at (1.75,0.7) {b};
\draw[fill=node_inner, draw=node_outer] plot[smooth cycle] coordinates {(4.5, 0) (5.5, 0.5) (6, 1) (5.5, 2) (4.5, 1.5)};
\draw[fill=node_inner, draw=node_outer] plot[smooth cycle] coordinates {(7, 0.25) (7.5, 0.5) (8, 0.75) (7.5, 1.5) (6.5, 1.5)};
\node[roundnode] (a) at (5.15,1) {};
\node[roundnode] (b) at (7.2,1) {};
\draw (5.15,1) -- (7.2,1);
\draw (4,0) -- (4,2.5);
\node (a) at (0,2.25) {A)};
\node (b) at (4.5,2.25) {B)};
\end{tikzpicture}}
    \caption{Edge positioning at two-dimensional nodes. (A) Edges may be placed at any location on a node, enabling multiple connections between the same pair of nodes. (B) Nodes are connected at their geometric center.}
    \label{fig:dimensionality}
\end{figure}
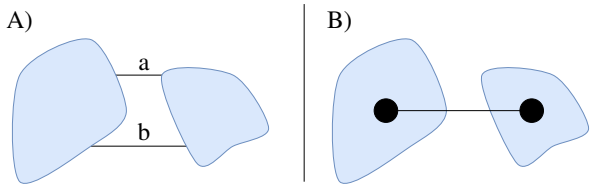

Spatial knowledge graph modeling enables the extension of nodes from zero-dimensional points (or edges from one-dimensional lines), into two-dimensional polygons, e.g. to represent spatial areas like regions or streets with different widths. However, this research did not identify any transformations of one-dimensional edges into two-dimensional representations. Several models, though, have been proposed for representing two-dimensional nodes. A key modeling decision in this context involves determining where edges can be positioned on two-dimensional nodes. 
Edges may either be attached anywhere on the nodes, e.g. Fall et al. \cite{Fall_2007} (Figure \ref{fig:dimensionality} A) or restricted to the geometric center of a two-dimensional node, e.g. Keitt et al. \cite{keitt_1997} (Figure \ref{fig:dimensionality} B).


\begin{table*}[t]
\centering
\caption{Spatial knowledge graph models}
\vspace{-3mm}
\footnotesize{Focus: D - Application Domain, M - Generic Graph Model, A - Analysis/Algorithm}  
\bigskip

\label{tab:graph-models-spatial}
\setlength{\tabcolsep}{0.6em}
\renewcommand{\arraystretch}{1.25} 
\begin{tabular}{l|l|cc|cc|ccc|c|c|l}
\multicolumn{1}{c|}{\textbf{}} & \multicolumn{1}{c|}{\textbf{}} & \multicolumn{4}{c|}{\textbf{Edges}} & \multicolumn{3}{c|}{\textbf{\begin{tabular}[c]{@{}c@{}}Spatial \\ Annotation\end{tabular}}} & \multicolumn{1}{c|}{\textbf{}} & \multicolumn{1}{c|}{\textbf{}} & \multicolumn{1}{c}{\textbf{}} \\
\multicolumn{1}{c|}{\textbf{Paper}} & \multicolumn{1}{c|}{\textbf{Year}} & \multicolumn{1}{l}{\rot{Undirected}} & \multicolumn{1}{l|}{\rot{Directed}} & \multicolumn{1}{l}{\rot{Explicit}} & \multicolumn{1}{l|}{\rot{Inferred}} & \multicolumn{1}{l}{\rot{Node}} & \multicolumn{1}{l}{\rot{Edge}} & \multicolumn{1}{l|}{\rot{Graph}} & \multicolumn{1}{c|}{\textbf{\rot{\begin{tabular}[c]{@{}c@{}}Node \\ Dimensionality\end{tabular}}}} & \multicolumn{1}{c|}{\textbf{Focus}} & \multicolumn{1}{c}{\textbf{Application Domain}} \\ \hline
Keitt et al. \cite{keitt_1997} & 1997 & x & & & x & x & & & 2-dim. & D & landscape graphs \\ 
Buhl et al. \cite{Buhl_2006} & 2006 & x & & x & & x & & & 0-dim. & D & street networks \\
Crucitti et al. \cite{Crucitti_2006} & 2006 & x & & x & & x & & & 0-dim. & D & street networks \\
Fall et al. \cite{Fall_2007} & 2007 & x & x & & x & x & x & & 2-dim. & M & landscape graphs \\
Scellato et al. \cite{scellato_2010} & 2010 & x & x & x & & x & & & 0-dim. & D & social networks \\
Zhong et al. \cite{Zhong_2014} & 2014 & & x & x & & x & & & 2-dim. & D & urban networks \\
Wang et al. \cite{Wang_2016} & 2016 & & x & x & & x & & & 0-dim. & D & social networks \\
Hristova et al. \cite{Hristova_2016} & 2016 & x & & x & x & x & & x & 0-dim. & D & social networks \\
Fang et al. \cite{Fang_2017} & 2017 & x & & x & & x & & & 0-dim. & A & social networks \\
Wang et al. \cite{Wang_2020} & 2020 & & x & x & x & x & & & 0-dim. & D & urban networks \\ 
\end{tabular}
\end{table*}

\subsubsection{\textbf{Summary}}
Spatial knowledge graphs are modeled according to (1) the direction and inference of edges, (2) the annotation of spatial information and (3) the dimensionality of nodes, as summarized in Table \ref{tab:graph-models-spatial}.  
Each row contains one publication that is further categorized according to its focus and application domains, analogous to our scheme in Table \ref{tab:graph-models} on temporal knowledge graph models (Sec. \ref{sec:temporal_kg}). 
Most of the proposed spatial knowledge graph models use undirected edges, while directed edges or combinations of both appear less frequently. In most cases, the edges are explicitly provided in the underlying data, although some authors only use inferred edges or combine both types of edges. Spatial annotation is almost always applied at the node level, edge- or graph-level annotations occur only occasionally. The models predominantly employ zero-dimensional nodes, with only two approaches supporting higher-dimensional node representations. Unlike temporal graphs, the majority of publications on spatial knowledge graphs focuses on a research question within a specific application domain (D) by combining use-case-specific graph models with new analysis methods. Only a few exceptions propose more generic models (M) or introduce domain-independent analysis methods (A). 
The application domains of interest mainly include landscape graphs, social networks and urban networks.

\section{Spatio-temporal knowledge graph models}\label{sec:stkg}

\noindent
Spatio-temporal knowledge graphs combine the characteristics of temporal and spatial knowledge graphs described in the previous sections, i.e. they enable the modeling of connected entities that evolve over time and are located in space. 
Despite their proximity to many real-world complex systems graph-based spatio-temporal data models have been put aside in the 2000s \cite{Siabato_2018}.
Based on temporal \cite{G_ting_1994} and spatial \cite{tansel_93} databases, the relevance of combining both is already evident in numerous spatio-temporal databases and models \cite{PELEKIS_2004}. There are methods for analyzing spatio-temporal data, but they still struggle with challenges in data acquisition and validation \cite{Ansari_2019} and lack the integration of contextual information, e.g. through knowledge graphs \cite{Yang_2024}.
In the 2010s, Del Mondo et al. \cite{del_mondo_2010} were among the first to reintroduce spatio-temporal graphs as a significant research topic. Recently, STKGs are increasingly applied in various use cases including human mobility prediction \cite{Wang_2021}, retail sales forecasting \cite{Yang_2024}, traffic volume prediction \cite{Yang_2024}, ship communication \cite{Zhang_2023}, traffic flow forecasting \cite{Guo_2019}, activity location identification \cite{Li_2025}, fire detection \cite{Ferreira_2020}, human action recognition \cite{Yan_2018}, medical imaging \cite{Gadgil_2020, Lu_2021}, pedestrian behavior prediction \cite{Mohamed_2020}, air quality forecasting \cite{Qi_2019}, wind turbine condition monitoring \cite{Jin_2024_2} and favorite location discovery \cite{Xiong_2020}.

In the following, we will first propose a principal guideline to model STKGs and then discuss foundational as well as recent STKG approaches in more detail. 

\begin{figure}[t!]
    \centering
    \includegraphics[width=0.7\linewidth]{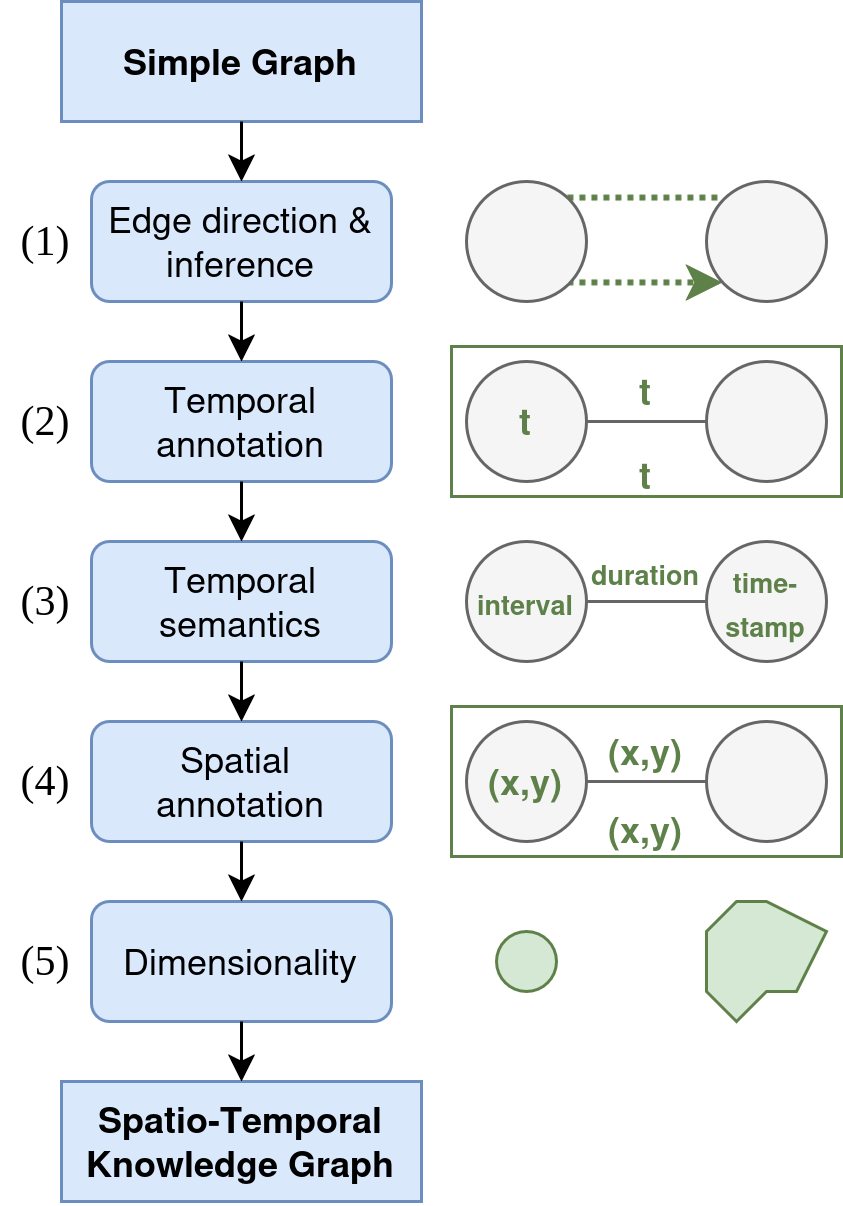}
    \caption{Modeling guideline: A spatio-temporal knowledge graph can be modeled by extending a simple graph model with additional properties and modifications that incorporate spatial and temporal information.}
    \label{fig:stkg_workflow}
\end{figure}

\subsubsection{\textbf{Modeling Guideline for Spatio-Temporal Knowledge Graphs}}

STKGs unify and extend the modeling principles of temporal and spatial knowledge graphs. We analyze STKG models according to the identified categories in sections \ref{sec:temporal_kg} and \ref{sec:spatial_kg} and provide a comprehensive categorization in Table \ref{tab:stkg-models}, which we summarize later in this section.
Based on our literature research and findings of the previous sections, we propose a general modeling guideline to support the development of STKGs in applications. The guideline is intended as a conceptual workflow rather than a prescriptive procedure: modeling decisions may be revisited iteratively, and their order is not fixed but depends on the specific requirements, data characteristics and objectives of the application domain.

During modeling \textbf{edge direction and inference} (see Section \ref{sec:temporal_kg} and \ref{sec:spatial_kg}) encompass determining if edges should be directed or undirected and whether they are explicitly defined or inferred through inference methods (Figure \ref{fig:stkg_workflow} (1)). If the semantic of edges depend on direction, directed edges are required, otherwise undirected edges may be used. When different edge semantics coexist within the same model both directed and undirected edges can be incorporated. Primary modeling approaches for spatial graphs have introduced various methods for edge inference in cases where edges are not known in advance. A model may include a combination of explicitly given and inferred edges.

The \textbf{temporal aspects} (see Section \ref{sec:temporal_kg}) of a spatio-temporal knowledge graph include the temporal annotation and the associated time semantic (Figure \ref{fig:stkg_workflow} (2, 3)). Temporal annotations can be applied at a node-, edge-, node-edge- or graph level. Node-annotated graphs store temporal information at the nodes and treat edges as static or externally defined, whereas edge-annotated graphs typically rely on an evolving set of edges and a static set of nodes. Node-edge-annotated graphs support temporal annotation at both nodes and edges. Graph-annotated approaches differ from the others in that they represent not a single graph but a collection of temporally annotated graphs.
The time semantic can be differentiated into timestamps, intervals and durations. Timestamps are typically represented as discrete integers, whereas intervals enable the modeling of continuous time. Durations usually appear in combination with one of the other temporal semantics.

The \textbf{spatial aspects} (see Section \ref{sec:spatial_kg}) of a spatio-temporal knowledge graph model are characterized by the spatial annotation and the dimensionality (Figure \ref{fig:stkg_workflow} (4, 5)). Spatial annotations can be applied at a node-, node-edge- or graph-level. Node-annotated graphs store the spatial information at the nodes. Node-edge-annotated graphs additionally store spatial information at the edges. Graph-annotated approaches typically consist of two separate graphs, one graph stores the spatial information at the nodes and another graph stores nodes without spatial information. Both graphs are connected via edges that link nodes without spatial information to nodes that contain spatial information. Spatial information can be stored as geometries, allowing nodes to be modeled in two or more dimensions.

In summary, a spatio-temporal graph model can be modeled by: (1) determine how edges are formed (explicit vs. inferred, directed vs. undirected), (2) decide where to attach temporal information (nodes, edges or graph-level) and (3) spatial information (nodes, edges or graph-level), (4) choose the temporal semantics (timestamp, intervals or duration) and (5) decide on the spatial dimensionality of nodes (points vs. regions).


\begin{figure}[b]
    \centering
    \includegraphics[width=1\linewidth]{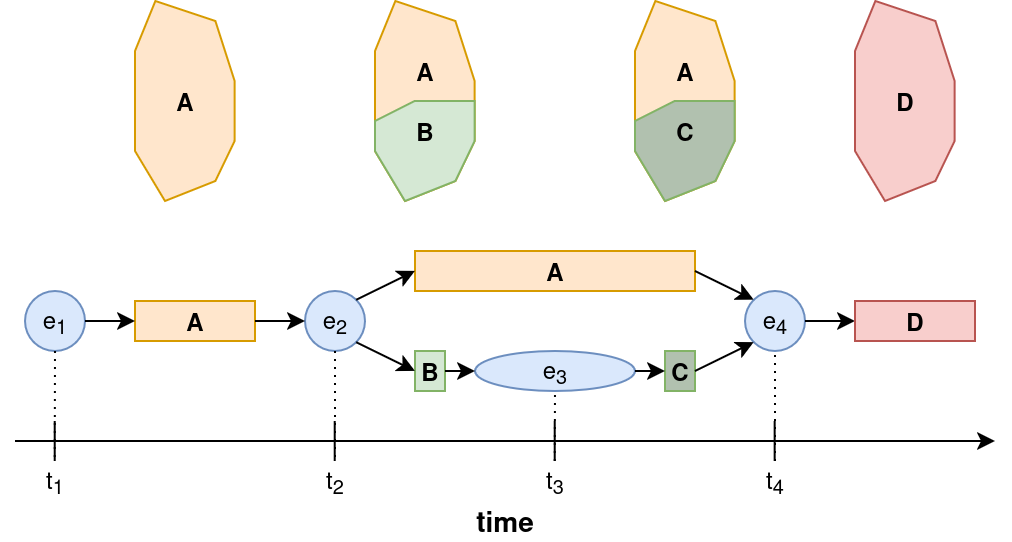}
    \caption{An example of a history graph according to Renolen \cite{renolen_1999}. The rectangles represent states with different duration, the circles show instantaneous event and ovals show longer-lasting events. The polygons illustrate a region and its different states according to the history graph.}
    \label{fig:history_graph}
\end{figure}

\subsubsection{\textbf{Foundational spatio-temporal graph models}}
\noindent
Renolen \cite{renolen_1999} proposed one of the first spatio-temporal graph models in 1996, called the ``History Graph Model''.
The model is conceptual, addressing only the basic principles of handling spatio-temporal data in a graph structure and does not include a formal definition or implementation.
The author states that changes, also referred to as events, on spatio-temporal objects can occur on several levels, e.g. attribute, geometrical or topological. Those events, e.g. a new street opened, can also be the result of longer processes, e.g. construction of a street. The author further describes the difference between state-based and event-based modeling approaches. In a state-based approach the history of an object can be seen as a sequence of states while in an event-based approach the history is viewed as a sequence of events. The reason for state changes get lost in state-based approaches. In event-based approaches the changes and states can be obtained by summing up the changes from an initial state but they have the disadvantage that the information for any given status must always be recalculated. The author argues that both models complement each other and combines them into the ``History Graph Model''. The model distinguishes between seven different types of events: 
Creation, Alteration, Destruction, Reincarnation, Split, Merge and Reallocation. Figure \ref{fig:history_graph} illustrates a history graph. A circle denotes an event and the width corresponds to the event duration. The possible events are based on the mentioned seven types, e.g. creation $e_1$ at $t_1$ and split $e_2$ at $t_2$. The rectangles present states and widths their valid interval. 
The polygons visualize the history graph as a region affected by events that lead to different states.

\begin{figure}[t]
    \centering
    \includegraphics[width=1\linewidth]{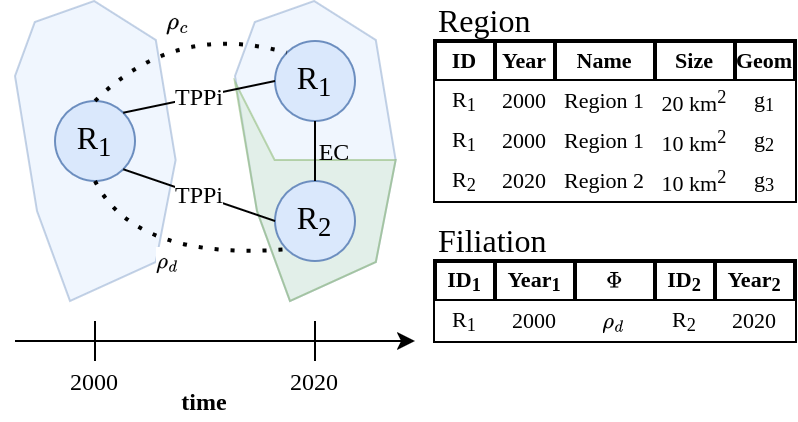}
    \caption{An example of a spatio-temporal knowledge graph (left) and its implementation in a relational database (right) according to Del Mondo et al. \cite{Del_Mondo_2013}. The graph shows regions over time together with their topological and filiation relations. The tables show the implementation of the graph on left in a relational database.}
    \label{fig:model_stkg_landscape}
\end{figure}

Del Mondo et al. \cite{Del_Mondo_2013} proposed one of the first formal definitions of spatio-temporal knowledge graphs and demonstrate principles of the model on the evolution of the regions and provinces of Chile. The authors define a spatio-temporal knowledge graph as $G=(V, E_{\Gamma}, E_{\Phi})$. 
$V$ is a set of vertices $(e,t)$ where $e$ is a label, e.g. a region and $t$ is an element of a list of time instances $T = (t_1,\ldots,t_n)$ with the same granularity.
$E_{\Gamma}$ is a set of tuples $((e_i, t_i), \gamma, (e_j,t_j))$ where $t_i \leq t_j$ and $\gamma \in \Gamma$. The topological relations $\Gamma$ describe the relationships of entities in space and include relations such as ``$e_1$ is a part of $e_2$'', ``$e_1$ overlaps $e_2$'' or ``$e_1$ is equivalent with $e_2$'' where $e_1,e_2 \in V$, they were surveyed for instance by Randell et al. \cite{randell_92} and Cohn et al. \cite{Cohn_1997}. 
$E_\Phi$ is a set of tuples $((e_i, t_i), \rho, (e_j,t_j))$ where $t_i \leq t_j$ and $\rho \in \Phi$. The filiation relations $\Phi$ relates to the identity of entities and can be divided into continuation $\rho_c$ and derivation $\rho_d$. A continuation relation $\rho_c$ exists if an entity $e$ continues to exist between $t_i$ and $t_j$, e.g. a region has the same name in two consecutive time instances. A derivation relation $\rho_d$ exists if an entity derives into other entities, e.g. a region splits into two regions with different names. 
In addition the authors describe the implementation of the spatio-temporal graph in a relational database. 
Entities are stored in entity tables $E(ID, Time, U_1,...,U_n,G)$ where $ID$ is the identifier of a node, $Time$ is the corresponding time instance $t \in T$, ($U_1,...,U_n$) is a possibly empty set of describing attributes and $G$ is a geometric attribute. A filiation table is defined as $F(ID_1, Time_1, \Phi, ID_2, Time_2)$ where $ID_1$ and $ID_2$ are the identifier of the connected nodes, $Time_1$ and $Time_2$ the corresponding time instances of the nodes and $\Phi$ is the derivation relation $\rho_d$. Continuation relations $\rho_c$ and topological relations $\Gamma$ can be derived from the data and therefore are not explicitly stored in the tables. 

Figure \ref{fig:model_stkg_landscape} illustrates the proposed spatio-temporal graph model (left) and the implementation in a relational database (right). Region $R_1$ splits between the years 2000 and 2020 into two regions $R_1$ and $R_2$ which are externally connected (EC) by a boundary. Both regions are an inverse tangential proper part (TPPi) of $R_1$, i.e. $R_1$ completely encompasses both regions and they share the same outer border. 
$R_1$ continued to exist (edge $\rho_c$) whereas $R_2$ forms a new region (edge $\rho_d$) derived from $R1$. The entity table $Region$ stores the regions together with their corresponding time instance, describing attributes and their geometry, e.g. their polygon. The filiation table stores the derivation filiation $\rho_d$ from $R_1$ to $R_2$. 


\begin{figure}[b!]
    \centering
    \includegraphics[width=0.9\linewidth]{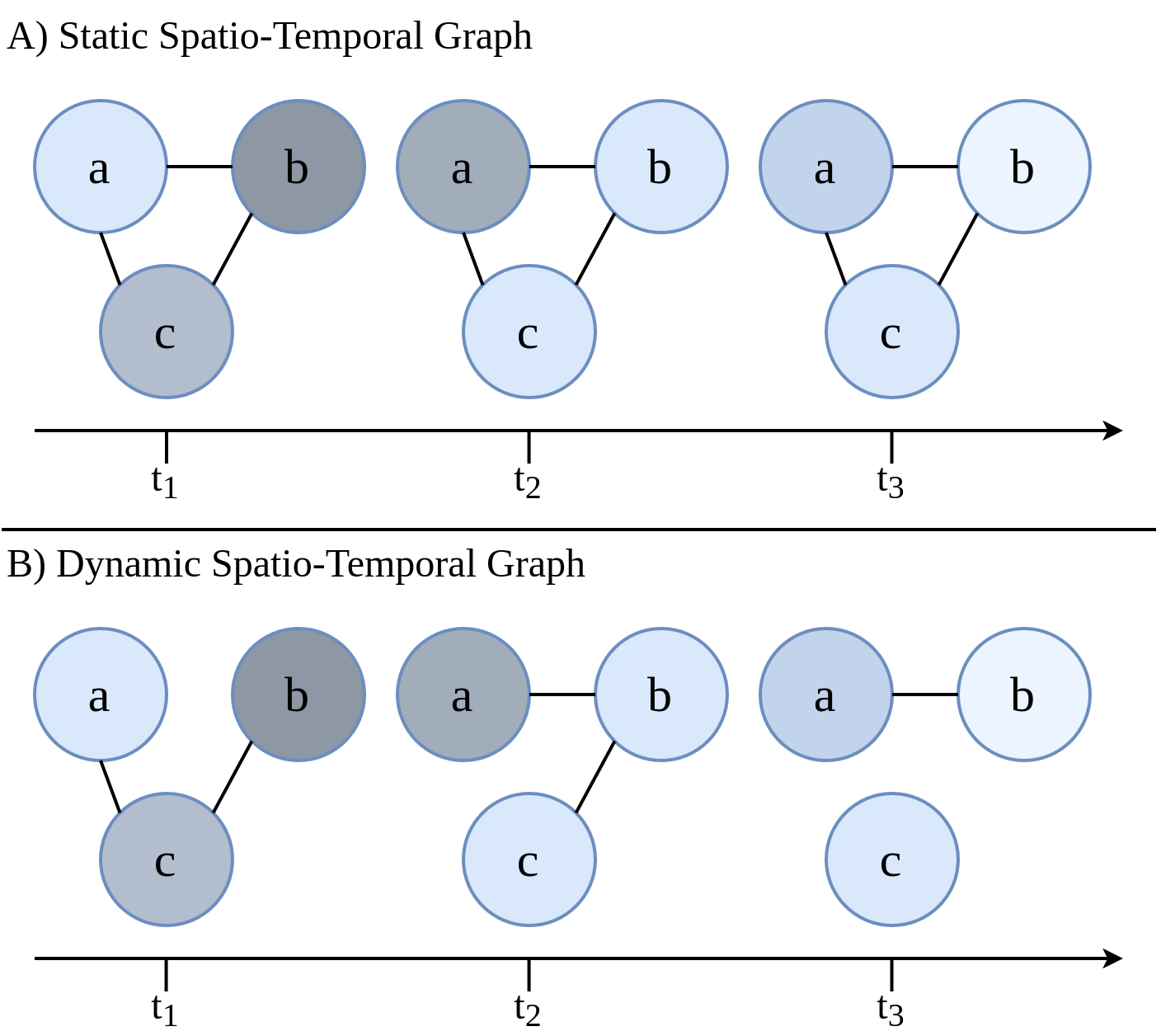}
    \caption{Types of spatio-temporal graphs according to Jin et al. \cite{Jin_2024}. A static spatio-temporal graph (A) has an invariant structure but the properties at the nodes can change. In a dynamic spatio-temporal graph (B) the structure and the properties can change.}
    \label{fig:static_dynamic_stg}
\end{figure}

\begin{table*}[t]
\centering
\caption{spatio-temporal knowledge graph models}
\vspace{-3mm}
\footnotesize{Focus: D - Application Domain, M - Generic Graph Model, A - Analysis/Algorithm}  
\bigskip

\label{tab:stkg-models}
\setlength{\tabcolsep}{0.5em}
\renewcommand{\arraystretch}{1.25} 
\begin{tabular}{l|l|cc|cc|ccc|ccc|ccc|c|c|l}
\multicolumn{1}{c|}{\textbf{}} & \multicolumn{1}{c|}{\textbf{}} & \multicolumn{4}{c|}{\textbf{Edges}} & \multicolumn{3}{c|}{\textbf{\begin{tabular}[c]{@{}c@{}}Temporal \\ Annotation\end{tabular}}} & \multicolumn{3}{c|}{\textbf{\begin{tabular}[c]{@{}c@{}}Temporal \\ Semantic\end{tabular}}} & \multicolumn{3}{c|}{\textbf{\begin{tabular}[c]{@{}c@{}}Spatial \\ Annotation\end{tabular}}} & \multicolumn{1}{c|}{\textbf{}} & \multicolumn{1}{c|}{\textbf{}} \\
\multicolumn{1}{c|}{\textbf{Paper}} & \multicolumn{1}{c|}{\textbf{Year}} & \multicolumn{1}{l}{\rot{Undirected}} & \multicolumn{1}{l|}{\rot{Directed}} & \multicolumn{1}{l}{\rot{Explicit}} & \multicolumn{1}{l|}{\rot{Inferred}} & \multicolumn{1}{l}{\rot{Node}} & \multicolumn{1}{l}{\rot{Edge}} & \multicolumn{1}{l|}{\rot{Graph}} & \multicolumn{1}{l}{\rot{Duration}} & \multicolumn{1}{l}{\rot{Interval}} & \multicolumn{1}{l|}{\rot{Timestamp}} & \multicolumn{1}{l}{\rot{Node}} & \multicolumn{1}{l}{\rot{Edge}} & \multicolumn{1}{l|}{\rot{Graph}} & \multicolumn{1}{c|}{\textbf{\rot{\begin{tabular}[c]{@{}c@{}}Node \\ Dimensionality\end{tabular}}}} & \multicolumn{1}{c|}{\textbf{Focus}} & \multicolumn{1}{c}{\textbf{Application Domain}} \\ \hline
Del Mondo et al. \cite{Del_Mondo_2013} & 2013 & x & x & x & x & x & & & & & x & x & & & 2-dim. & M & cadastral graphs \\
Williams and Musolesi \cite{Williams_2016} & 2016 & & x & x & & & & x & & & x & x & & & 0-dim. & M & urban, social, biological networks \\
Landesberger et al. \cite{von_Landesberger_2016} & 2016 & & x & x & & & x & & & & x & x & & & 0-dim. & A & user mobility networks  \\
Yan et al. \cite{Yan_2018} & 2018 & x & & x & & x & & & & & x & x & & & 0-dim. & D & skeleton graphs \\
Yu et al. \cite{Yu_2018} & 2018 & x & x & x & & & & x & & & x & x & & & 0-dim. & D & urban networks \\
Guo et al. \cite{Guo_2019} & 2019 & x & & x & & x & & & & & x & x & & & 0-dim. & D & street networks \\
Ferreira et al. \cite{Ferreira_2020} & 2020 & x & x & x & x & x & & & & & x & x & & & 0-dim. & M & fire detection networks \\
Mohamed et al. \cite{Mohamed_2020} & 2020 & x & & & x & & & x & & & x & x & & & 0-dim. & D & human trajectory networks \\
Zhang et al. \cite{Zhang_2020} & 2020 & & x & x & & & & x & & & x & x & & & 2-dim. & A & urban networks \\
Gadgil et al. \cite{Gadgil_2020} & 2020 & x & & & x & x & & & & & x & x & & & 0-dim. & D & brain networks \\
Lu et al. \cite{Lu_2021_2} & 2021 & x & & & x & x & & & & & x & x & & & 0-dim. & D & medical networks \\
Li et al. \cite{Li_2025} & 2025 & & x & x & x & x & & & & & x & x & & & 2-dim. & D & mobile phone activity networks \\
\end{tabular}
\end{table*}

\subsubsection{\textbf{Contemporary approaches and applications}}

\noindent
Subsequent to the early foundational models, a wide range of new models has emerged over time, often tailored to specific applications or analysis methods. Jin et al. \cite{Jin_2024} observed an increasing trend of publications related to the topic of spatio-temporal graph neural networks (STGNN) in their 2024 survey. The authors categorize the primary application domains in transportation, environment, public safety and public health. 
In their observations, a spatio-temporal knowledge graph can often be defined as $G_t = (V, E_t, A_t)$. $V$ is a mostly invariant set of vertices, $E_t$ is a possibly time-varying set of edges at time $t$ and $A_t$ is the adjacency matrix at time $t$ and changes with $E_t$ respectively. The edges can either be directed or undirected and weighted or unweighted. The adjacency matrix can be created according to various rules, which are generally categorized as topology-based, distance-based, similarity-based and interaction-based.
Authors mainly focus on observations obtained by sensors, therefore they define the observations $X = \{X_t \in \mathbb{R}^{N \times F}| t=0, \ldots,T\}$ where $N$ are the spatial vertices and $F$ are the features. 
Depending on the task and data, two types of spatio-temporal graphs can applied as illustrated in Figure \ref{fig:static_dynamic_stg}. In a static spatio-temporal graph (Figure \ref{fig:static_dynamic_stg} A) the properties of nodes can change but the edges are stable. In a dynamic spatio-temporal graph (Figure \ref{fig:static_dynamic_stg} B) the edges as well as the properties of the nodes can change. 

Recently, Zeghina et al. \cite{Zeghina_2024} conducted a systematic analysis of spatio-temporal graphs
for deep learning approaches. 
The authors define an often applied STKG as a 4-tuple $G=(V,E,T,X)$, where $V$ is a set of vertices representing spatial entities, $E$ is a set of edges and $T$ is a set of time intervals or timestamps. $X = \{x_1,\ldots,x_n\} \cup \{y_1,\ldots,y_m\}$ is a set of properties associated with the vertices and edges, where $x_i$ represents properties of vertex $v_i$ and $y_i$ represents the properties of edge $e_i$. They discovered that most use cases of spatio-temporal graphs avoid sophisticated modeling and prefer to use simple graph models, like the one proposed by Jin et al. \cite{Jin_2024}. One major drawback of typical STKGs 
is that they are application-specific and cannot be applied beyond the scope of their initial context.


\subsubsection{\textbf{Summary}}
Spatio-temporal knowledge graphs combine the already complex concepts of temporal and spatial graphs. Foundational models demonstrate that incorporating time and space significantly increases both the complexity and the amount of modeling required to represent real-world systems. In contrast, many recent approaches deliberately reduce this complexity by adopting simplified graph structures that omit rich temporal or spatial semantics. While such simplifications often facilitate analysis or learning tasks, they typically result in models that are tightly coupled to a specific application context and difficult to reuse beyond it. Table \ref{tab:stkg-models} summarizes how spatio-temporal knowledge graphs are modeled in the literature, based on the categorization schemes introduced for temporal (Table \ref{tab:graph-models}) and spatial (Table \ref{tab:graph-models-spatial}) graph models. The comparison covers key modeling dimensions including year, edge direction and inference, temporal annotation and semantics, spatial annotation, node dimensionality, research focus, and application domains.
Overall, the number of models using directed and undirected edges is comparable, and several approaches support mixed edge types within a single graph. Most models rely on explicitly defined edges, while some additionally incorporate inference mechanisms or combine explicit and inferred relationships. Temporal information is most frequently attached at the node level, followed by graph-level annotation, while all surveyed models adopt timestamps as their primary temporal semantic. Similarly, spatial information is consistently modeled at the node level across all approaches. The majority of STKGs employ zero-dimensional nodes, with only a limited number of models supporting higher-dimensional spatial representations such as regions.
With respect to research focus, most publications address domain-specific research questions (D), whereas comparatively few emphasize reusable generic models (M) or standalone analysis methods (A). Application domains are diverse, with urban networks, biological systems, and human mobility being among the most frequently studied.
In summary, STKGs extend temporal and spatial knowledge graphs by incorporating the respective missing dimension, thereby enabling more expressive modeling and analysis of complex spatio-temporal phenomena. However, the complexity of the proposed graph models strongly depends on the application domains and objectives. This heterogeneity highlights the need for more systematic modeling guidelines and unified abstractions, and it gives rise to a set of open challenges that are discussed in the following section.

\section{Challenges}\label{sec:challenge}

\noindent
Despite increasing interest and progress, numerous open questions, challenges and future research directions remain when working with spatio-temporal knowledge graphs. The following challenges span the entire life cycle of such graphs, from modeling and construction to maintenance, analysis and reuse. The list is not exhaustive, as the field is still evolving and many issues are closely interrelated. In particular, several challenges arise repeatedly across different stages and must be addressed jointly rather than in isolation.

\subsubsection{\textbf{Modeling}}
As shown in the previous sections, the landscape of modeling approaches and terminologies used for temporal, spatial and spatio-temporal knowledge graphs is very heterogeneous. Many of the spatio-temporal knowledge graph representations proposed in current research are tailored to specific applications rather than designed for use across domains, resulting in a wide variety of models and reduced transferability \cite{Zeghina_2024}. Some authors point out the lack of simple and effective methods for modeling such graphs or for using the knowledge they contain in a versatile way \cite{Yang_2024}. Additionally, many publications focus on urban networks leaving numerous research areas underexplored, such as agriculture, geology, sports, epidemiology or weather forecasting \cite{Zeghina_2024}. Developing generalizable models with broad cross-domain applicability remains an open challenge, which this work addresses and future studies may further explore. Spatio-temporal knowledge graphs should be regarded not just as practical tools, but as long-term structures for maintaining and preserving interconnected temporal and spatial data.

\subsubsection{\textbf{Construction}}
Spatio-temporal data within many domains is rarely available in graph form and is more commonly found in tabular formats or unstructured text. After deciding on suitable spatio-temporal knowledge graph models, methods could be developed to create the graphs \cite{Hofer_2024} and manage them throughout their life cycle \cite{geisler_2025}. 
Manual creation processes could be replaced by semi-automatic or fully-automatic methods, for instance by using Large Language Models (LLMs) to generate a spatio-temporal knowledge graph directly from unstructured \cite{Pan_2024} or tabular \cite{chen_2025} data.  A key challenge for knowledge graphs is that they often suffer from incompleteness caused by missing information in nodes or edges. Knowledge graph completion methods typically rely on expert knowledge \cite{Plamper_2023}, artificial intelligence \cite{Lin_2015, Chen_2020, cai_2022} or alternative inference methods \cite{Jin_2024}. The benefits of reliable (semi-)automated creation methods include increased speed and the ability to automatically recommend entities and relationships.

\subsubsection{\textbf{Provenance}}
Spatio-temporal data used to construct spatio-temporal knowledge graphs often originates from heterogeneous sources that differ in quality, granularity and reliability. To ensure trustworthiness and reproducibility, it is therefore essential to explicitly record the provenance of entities, relationships and their temporal and spatial annotations. Provenance information enables users to assess data origin, applied transformations, inference steps and underlying assumptions.
In spatio-temporal knowledge graphs, provenance is tightly coupled with temporal aspects, as data may evolve, be corrected retrospectively or become outdated. Without systematic provenance tracking, it becomes difficult to distinguish historical states from inferred or updated knowledge. Provenance is also crucial for downstream analyses, particularly when data is subject to legal, ethical or organizational constraints, such as protected locations or time-dependent access restrictions.
Despite its importance, provenance is often treated as auxiliary metadata rather than an integral modeling component. Developing approaches that tightly integrate provenance with temporal and spatial information remains an open challenge for enabling reliable interpretation, auditing and long-term maintenance of spatio-temporal knowledge graphs.

\subsubsection{\textbf{Integration}}
A spatio-temporal knowledge graph should preserve interconnected temporal and spatial data over the long term rather than being constructed for each individual use case. Consequently, new information, such as knowledge graphs from other projects, may need to be linked to an existing spatio-temporal knowledge graph. 
Combining two knowledge graphs that employ different temporal and spatial annotations or semantics can lead to significant issues, for instance when merging different types of temporal or spatial annotation.
Therefore, the process of data integration should be carefully planned and managed in advance to ensure the long-term preservation of knowledge within a spatio-temporal knowledge graph while maintaining its full functionality. 

\subsubsection{\textbf{Incremental updates}}
Spatio-temporal knowledge graphs are inherently dynamic, as the real-world systems they represent evolve over time. Temporal changes may affect the attributes of nodes and edges, the existence of entities and relationships or the associated spatial information. To remain useful, such changes must be incorporated incrementally into the knowledge graph while preserving internal consistency and historical traceability \cite{Hofer_2024, Gross_2016}.
Incremental updates introduce several challenges. Updates must be modeled in a way that allows past states of the graph to be reconstructed, enabling longitudinal analyses and retrospective reasoning. Moreover, updates may originate from heterogeneous sources with different update frequencies, levels of reliability or temporal granularities, complicating synchronization and conflict resolution. Incremental updates may also interact with inferred knowledge, requiring mechanisms to revise or invalidate previously derived relationships when new evidence becomes available.
Many existing approaches treat spatio-temporal knowledge graphs as static snapshots or periodically rebuilt structures, which limits their applicability for long-running and evolving systems. Developing methods for efficient, consistent and provenance-aware incremental updates remains a key challenge for enabling long-term maintenance, scalability and sustainability of spatio-temporal knowledge graphs.

\subsubsection{\textbf{Ethics}}
Spatio-temporal knowledge graphs should serve as a form of knowledge representation that is reusable and accessible to as many users as possible, while respecting potential legal, organizational and ethical constraints. 
Although access to all data cannot always be granted to every user, following the FAIR principles (Findable, Accessible, Interoperable, Reusable) \cite{Wilkinson_2016} offers benefits for both public use cases, e.g. open source projects or private use cases in enterprise applications. 
In this sense, spatio-temporal knowledge graphs could be extended with metadata that capture the different modeling approaches described in the previous sections, making it easier to identify and integrate compatible knowledge graphs and thereby supporting reusability.
When modeling large spatio-temporal knowledge graphs that integrate heterogeneous and possibly sensitive data, the CARE principles (Collective benefit, Authority to control, Responsibility, Ethics) \cite{Carroll_2020} should be considered to account for the diverse actors involved. 
Future research can contribute frameworks and guidelines to maintain communication with all involved parties throughout the modeling and development process to avoid impacts, such as revealing protected locations, disobeying data authorities or overlooking bias triggering wrong predictions. 

\subsubsection{\textbf{Evaluation}}
The Evaluation of spatio-temporal knowledge graphs is essential and should be viewed as a continuous process. It must assess not only the quality of the initial graph construction but also the quality of incremental updates and various integration tasks. Moreover, the quality of spatio-temporal knowledge graphs directly influence the results of subsequent analyses. Currently, there is no universal gold standard or ground truth for knowledge graphs and quality checks often need to be performed manually by experts \cite{Qudus_2025, Qudus_2023}. 
Evaluation challenges arise even with smaller knowledge graphs and become increasingly complex for large, complex and automatically generated spatio-temporal knowledge graphs. As a result, knowledge graph quality management is emerging as an important topic, similar to the long-established practices in relational data management \cite{Xue_2022}. 
The creation of benchmarks for spatio-temporal knowledge graphs is challenging, as it requires accounting for multiple and sometimes unknown quality criteria. Additionally, inconsistencies in temporal and spatial properties make evaluating overall quality more complex. Consequently, many existing approaches to spatio-temporal knowledge graphs therefore rely on synthetic data or on easy accessible sources \cite{Wang_2021}, but both performance measurements on real-world data sets and suitable evaluation methods are missing \cite{Li_2025, Zhang_2023}.

\subsubsection{\textbf{Analyses}}
The long history of graph theory has produced a wide variety of analysis methods for static, temporal and spatial graphs, but most of these methods do not transfer directly to spatio-temporal knowledge graphs.
Due to their heterogeneous and complex nature, spatio-temporal knowledge graphs are difficult to analyze, as many analysis methods still assume homogeneous graph structures.
Most research on spatio-temporal knowledge graphs focuses on forecasting within deep learning tasks, possibly overlooking traditional graph analysis tasks such as clustering, motif discovery, anomaly detection or recommendation \cite{Zeghina_2024}.
Current analysis methods often build on Graph Neural Networks (GNNs) and enhance performance through transforming the graph structure into a low-dimensional vector space \cite{Grover_2016} or use case specific model design, yet they lack interpretability, making it difficult to establish trust in their capabilities among decision makers \cite{Jin_2024}.

\subsubsection{\textbf{Visualization}}
A key strength of graphs is their suitability for visualization, due to their pictorial nature. But visualization becomes considerably more challenging as graph models grow in complexity, particularly when visualizing additional properties such as time \cite{Beck_2016} and space \cite{hadlak_2015}, as in spatio-temporal knowledge graphs.
Major challenges include visualizing temporal changes, mapping nodes and edges onto geographic representations and navigating within the graph across time and space. 
While the spatio-temporal knowledge graph could be maintained purely as a data structure, the actual graph itself is often of interest for analysis.
In a recent survey Deng et al. \cite{Deng_2023} examined current developments and trends of spatio-temporal visualizations, covering both general approaches and those specifically applied to graph structures.

\section{Conclusion}\label{sec:conclusion}
\noindent
Spatio-temporal knowledge graphs provide a powerful means to represent temporal and spatial information within a unified graph structure and have emerged as an important research domain for modeling complex real-world systems that inherently rely on spatio-temporal data. Their modeling requires the integration of principles from both temporal and spatial knowledge graph models, which have evolved independently across different research communities. As a result, rapid development, diverse application areas and heterogeneous contributions have led to inconsistent terminology and modeling decisions, complicating literature discovery and hindering standardization.

This survey has shown that many existing spatio-temporal knowledge graph models and analysis methods are tailored to specific use cases, which limits their reusability, long-term knowledge preservation and cross-domain applicability. The proposed modeling guideline aims to support researchers and practitioners in adopting a more systematic and transparent approach to spatio-temporal knowledge graph design. A stronger focus on generalizability, reusability and long-term durability could enable the development of persistent spatio-temporal knowledge graphs that preserve not only current information but also its evolution and provenance. In the long term, linking and integrating multiple spatio-temporal knowledge graphs may facilitate broader analyses and enable new cross-domain insights.

\section*{Acknowledgments}
\noindent
Funded by the Deutsche Forschungsgemeinschaft (DFG, German Research Foundation) Project-ID 528485254 - FIP 16 and by the Europäischer Fonds für regionale Entwicklung (EFRE, European Regional Development Fund) Project number ZS/2023/12/182018. 

\bibliographystyle{IEEEtran}
\bibliography{references_survey.bib}

\section{Biography Section}



\vspace{-25pt}

\begin{IEEEbiographynophoto}{Philipp Plamper} 
    working toward the PhD degree in computer science at Anhalt University of Applied Sciences in Köthen (Germany). His research interests include knowledge graphs as well as data engineering and management.
\end{IEEEbiographynophoto}

\vspace{-25pt}

\begin{IEEEbiographynophoto}{Hanna Köpcke}
    is currently a professor at University of Applied Sciences Mittweida (Germany). Her research interests include data management and data integration, with a particular focus on entity resolution, knowledge graphs, and information retrieval.
\end{IEEEbiographynophoto}

\vspace{-25pt}

\begin{IEEEbiographynophoto}{Anika Groß}
  is currently a professor at Anhalt University of Applied Sciences in Köthen (Germany). Her research focuses on data integration, including the linking and clustering of knowledge sources, particularly knowledge graphs and ontologies, as well as their evolution and analysis.

\end{IEEEbiographynophoto}

\vfill

\end{document}